\documentclass[12pt,amsmath,amssymb,floatfix]{revtex4}
\usepackage{graphicx}

\begin{document}

\title{Nonequilibrium study of the  $J_{1}-J_{2}$ Ising model with random $J_{2}$ couplings in the square lattice}

\author{Octavio D. Rodriguez Salmon}
\affiliation{Departamento de F\'{\i}sica, Universidade Federal do Amazonas, 3000, Japiim, 69077-000, Manaus-AM, Brazil}
\email{octaviodrs@ufam.edu.br}
\author{Minos A. Neto}
\affiliation{Departamento de F\'{\i}sica, Universidade Federal do Amazonas, 3000, Japiim, 69077-000, Manaus-AM, Brazil}
\author{Thiago Lobo}
\affiliation{Instituto Federal de Educa\c{c}\~ao, Ci\^{e}ncia e Tecnologia do Rio de Janeiro - IFRJ\\ Rua Dr. Jos\'e Augusto Pereira dos Santos, s/n (C.I.E.P. 436 Neusa Brizola), 24425-004 S\~ ao Gon\c{c}alo, Rio de Janeiro, Brasil.}
\author{Francisco Din\'{o}la Neto} 
\affiliation{Centro Universit\'{a}rio do Norte - UNINORTE\\Rua Huascar de Figueiredo, 290, Centro\\ 69020-220, Manaus, AM, Brazil.}

\begin{abstract}
We studied the critical behavior of the $J_{1}-J_{2}$ spin-{1/2} Ising model in the square lattice by considering $J_{1}$ fixed and $J_{2}$ as random interactions following discrete and continuous probability distribution functions. The configuration of  $J_{2}$ in the lattice evolves in time through a competing kinetics using Monte Carlo simulations  leading to a steady state without reaching the free-energy minimization. However, the resulting non-equilibrium phase diagrams  are, in general, qualitatively similar to those obtained   with quenched randomness at equilibrium in past works. Accordingly, through this dynamics the essential critical behavior at finite temperatures can be grasped for this model. The  advantage is that simulations spend less computational resources, since the system does not need to be replicated or equilibrated with Parallel Tempering. A special attention was given for the value of the amplitude of the  correlation length  at the critical point of the superantiferromagnetic-paramagnetic transition. 
\end{abstract}

\maketitle

\section{Introduction}

Nonequilibrium classical spin Hamiltonians with random parameters are still attracting the studies of Statistical Mechanics \cite{mijat}. Problems like spin glasses were generally studied by  spin models with quenched randomness, where the couplings between pairs of spins are fixed in time, but their  values are randomly distributed throughout the lattice. So this cause frustration, which is a fundamental ingredient for this glassy state. In this context the model is studied at equilibrium. Nevertheless, this neglects the possibility that  frustration could avoid the system to reach an equilibrium  state and so the macroscopic behavior would be determined by the dynamics. 

In order to simulate this situation, a competing kinetic process was taken into account to induce the presence of nonequilibrium steady states \cite{garrido_1991}.  On the other hand, quenched randomness disregard diffusion of magnetic ions  \cite{lopez_1992}. When diffusion occurs, the distance between pairs of spins varies with time. A way of modeling this phenomena in a lattice, consists in considering the exchange couplings $J_{ij}$ to change in time and space \cite{gonzalez_1994}. Past studies  reported that the diffusion of disorder affect the macroscopic behavior of the system \cite{gonzalez_1994}. \smallbreak

A kind of conflicting dynamics was implemented  to study   the random-field Ising model (RFIM) \cite{nuno_2009, nuno_2010} inspired in the work of Gonzalez et. al. \cite{gonzalez_1994}.  Interestingly, it has been demonstrated \cite{paula_2014} that  quenched randomness destroys any first-order phase transition in low dimensions. Thus, the RFIM does not present a tricritical point in the ferromagnetic-paramagnetic frontier bellow the upper critical dimension ($d_{c} = 6$) \cite{hartman_2011}. However, in the context of this competing kinetic, the phase diagram of the RFIM  exhibits a tricritical point when the field obeys  a double-Gaussian probability distribution in the square lattice \cite{nuno_2009}, and for a bimodal distribution in the cubic lattice \cite{nuno_2010}. Accordingly, these facts motivate the comparison between results with quenched randomness and this conficting dynamics. \smallbreak

 In this paper we studied the $J_{1}-J_{2}$ Ising Model with in the particular case where the next-nearest-neighbor interactions $J_{2}$ are random, and evolve in time by a competing kinetics which is detailed in the next section \cite{nuno_2009,nuno_2010}. This model with fixed $J_{1}$ and $J_{2}$ antiferromagnetic interactions has shown controversial results in its critical behavior \cite{nightingale_1977,swendsen_1979, otimaa_1981,binder_1980,landau_1980,landau_1985,landau_2000}. Its  phase diagram is studied in the plane where the horizontal axis is either   $r = |J_{1}/J_{2}|$ or  $r = |J_{2}/J_{1}|$ , and the vertical axis is the temperature $T$. For $J_{2} < J_{1}$ the low temperature order is antiferromagnetic \textbf{AF}, and for $J_{2} < J_{1}$ the colinear or superantiferromagnetic \textbf{SAF}  order is present. It was believed that for $  J_{1}/2 < J_{2} \lesssim J_{1}$ the critical frontier  was continuous with weak universality, i.e. exponents like $\beta$ and $\gamma$ break the universality, but their normalized exponents $ \beta/\nu$ and $\gamma/\nu$ remain the same. However, a later work showed a double-peaked structure of the  energy histogram suggesting weak first-order transition \cite{kalz_2011}. 

When $J_{2}$ takes the value $\lambda_{1} J_{1}$ with probability $p$ and $-\lambda_{2} J_{1}$ with probability $1-p$, a competition emerges between  \textbf{SAF} and \textbf{AF} phases if $J_{1}$ is antiferromagnetic, and  between \textbf{SAF} and \textbf{F} if $J_{1}$ is ferromagnetic (with $\lambda_{1}  > 0$ and $\lambda_{2} \geq 0$). These two cases were studied considering quenched randomness. In the first one  the authors simulated the model with bond dilution in $J_{2}$ ($\lambda_{1} = 1$ and $\lambda_{2} =0$) \cite{yining_2018}. They used Parallel Tempering so as to equilibrate the system. This Monte Carlo technique is commonly used for equilibrate  models with quenched randomness at low temperatures. So it  must be executed in parallel  for simulating each replica of the system at different temperatures for each bond realization \cite{swendsen_1986,david_2005}. 

The authors \cite{yining_2018} obtained a phase diagram in the plane $T$ versus the degree of dilution $x=p$, where the frontiers \textbf{SAF-P} and \textbf{AF-P} are separated by a gap between $0.3 < x < 0.8 $ at $T=0$. They found strong evidences of a spin-glass phase in this gap only at zero temperature, regarding that there should not be spin-glass phase  in two dimensions for Ising-type spin-glass models (see references \cite{nobre_2001} and \cite{fernandez_2016} for this controversy). On the other hand, the second case with $\lambda_{1} = \lambda_{2} =1$ was treated in the effective-field approximation obtaining a phase diagram with similar topological characteristics \cite{octavio_2009}, though the nature of the gap at $T=0$ remains an open question. So, in this work we mainly simulate these two cases in which the random bonds $J_{2}$ change in time by a conflicting dynamics \cite{nuno_2009, nuno_2010}. 


\section{The Model and Simulation}

The system is modeled by the following Hamiltonian:

\begin{equation}
 \displaystyle {\cal H}(\boldsymbol{S,J_{2}}) = -J_{1} \sum_{(nn)} S_{i}S_{j} - \sum_{(nnn)} J_{2}^{(i,j)} S_{i}S_{j} ,
\label{ham}
\end{equation}
\vskip \baselineskip 
\noindent
where $S_{i} = \pm 1$, for $i=1,2,3,...,N$, with $N=L\times L$ is the number of sites of a square lattice of size $L$.  $\mathbf{S}=\{S_{1},S_{2}, \dots  S_{N}\}$ is the set of spin variables, and $\mathbf{J_{2}} = \{ J_{2}^{(i,j)} \}$ is the set next-nearest-neighbor coupling values. 

In this work the following probability distributions functions (PDF) for $J_{2}$ are used:

\begin{equation}
\displaystyle {\cal P}(J_{2}^{(i,j)}) = p \delta(J_{2}^{(i,j)}-J_{1}) + (1-p) \delta(J_{2}^{(i,j)}+J_{1}),
\label{prob1}
\end{equation}%
where $ 0 \leq p \leq 1 $ and $J_{1} >0$ (ferromagnetic nearest-neighbor couplings). So by this \textbf{PDF}  each  coupling $J_{2}^{(i,j)}$ takes the value $J_{1}$ with probability $p$, and the value $-J_{1}$ with probability $1-p$. This is the case studied with quenched randomness in reference \cite{octavio_2009}. Then

\begin{equation}
\displaystyle {\cal P}(J_{2}^{(i,j)}) = p {\cal G}(J_{2}^{(i,j)}-J_{1}) + (1-p) {\cal G}(J_{2}^{(i,j)}+J_{1}),
\label{prob2}
\end{equation}%
where $J_{1} >0$ and the function is a Gaussian one:

\begin{equation}
\displaystyle {\cal G }(u) = \frac{1}{\sqrt{2\pi}\sigma} \exp(-\frac{u^{2}}{2 \sigma^{2}}) , 
\label{gaussiana} 
\end{equation}%
where $\sigma$ is the standard deviation. Then

\begin{equation}
\displaystyle {\cal P}(J_{2}^{(i,j)}) = p \delta(J_{2}^{(i,j)}) + (1-p) \delta(J_{2}^{(i,j)}-J_{1}),
\label{prob3}
\end{equation}%
where $J_{1} < 0$ (antiferromagnetic nearest-neighbor couplings). In this \textbf{PDF}  each  coupling $J_{2}^{(i,j)}$ is zero with probability $p$ (bond dilution), and  takes the value $J_{1}$ with probability $1-p$. This is the \textbf{PDF} studied with quenched randomness in reference \cite{yining_2018}. The last distribution used in this work is this double Gaussian \textbf{PDF} :

\begin{equation}
\displaystyle {\cal P}(J_{2}^{(i,j)}) = p{\cal G}(J_{2}^{(i,j)}) + (1-p) {\cal G}(J_{2}^{(i,j)}-J_{1}),
\label{prob4}
\end{equation}%
where $J_{1} < 0$. Note that for $\sigma \to 0$, we recover  the bimodal distribution given in Eq.(\ref{prob3}). The system is in contact with a thermal reservoir, which provides the temperature $T$ of the system inducing stochastic changes given by the following master equation:

\begin{equation}
 \displaystyle \frac{\partial P (t,\boldsymbol{S^{b}})}{\partial t} = \sum_{\boldsymbol{S^{a}}} ( W(\boldsymbol{S^{a}}\rightarrow \boldsymbol{S^{b}}) P(t,\boldsymbol{S^{a}}) - W(\boldsymbol{S^{b}} \rightarrow \boldsymbol{S^{a}}) P(t,\boldsymbol{S^{b}})   ), 
\label{mastereq}
\end{equation}%
being $W(\boldsymbol{S^{a}} \rightarrow \boldsymbol{S^{b}})$ the probability per unit time for a transition from the spin configuration $ \boldsymbol{S^{a}}$   to $\boldsymbol{S^{a}}$, and $P(t,\boldsymbol{S^{b}})$ the probability that the system is in the spin configuration $\boldsymbol{S^{b}}$ at time $t$. A stationary solution of the master equation is given by the detailed balance condition 
$ W(\boldsymbol{S^{b}} \rightarrow \boldsymbol{S^{a}}) = \exp(-\beta \Delta{\cal H})$, where $\Delta{\cal H} = {\cal H}(\boldsymbol{S^{a},J_{2}}) -{\cal H}(\boldsymbol{S^{b},J_{2}}) $, and $\beta = 1 / (k_{B}T)$. So, a steady state is reached by the Metropolis Algorithm $ W(\boldsymbol{S^{b}} \rightarrow \boldsymbol{S^{a}}) = \min\{1, \exp(-\beta \Delta{\cal H})\}$ \cite{gonzalez_1994,nuno_2009, nuno_2010}. 

To implement it in a simulation a Monte Carlo sweep (\textbf{MCS}) consists in generating a new configuration  of random $J_{2}$ bonds according to a given \textbf{PDF}; then, all lattice sites are visited, and each  spin  ﬂips  according to the Metropolis’ criterium. In this way the random bonds $J_{2}$ vary with time by changing their values after each \textbf{MCS}. This is the essence of this conflict dynamics where there is a competing kinetic between spins and random bonds. Thus the system reaches a  nonequilibrium steady state. This context  differs  from those of    equilibrium quenched and annealed random models.  
\smallbreak

For a given lattice size $L$ of the system,  we note that the equilibration is guaranteed after $10^{5}$ \textbf{MCS} for a given temperature. To overcome the critical slowing down we accumulated values each $L$ \textbf{MCS}. In some cases we run up to $3.6 \times 10^6$ \textbf{MCSs} to reduce  fluctuations in the thermal mean about the critical temperature. In this way the  order parameter, the susceptibility, the Binder Cumulant and the correlation length were averaged for a given temperature.  In this work we used three order parameters according to the phase formed by the parameters established in a particular  simulation. The first one is the magnetization per spin, which measures the ferromagnetic order  \textbf{F}:

\begin{equation}
\displaystyle m_{F} =  |\frac{1}{N} \sum_{i=1}^{N} S_{i} | . 
\label{mf}
\end{equation}

The antiferromagnetic order parameter meassuring the Neel order or the \textbf{AF} order is as follows :

\begin{equation}
\displaystyle m_{AF} = | \frac{1}{N} \sum_{i=1}^{N} (-1)^{x_{i}+y_{i}}S_{i} |. 
\label{maf}
\end{equation}

The order parameter for the colinear order (or the superantiferromagnetic order) is given by :

\begin{equation}
\begin{aligned}
\displaystyle  m_{SAF} &= \sqrt{m_{x}^{2}+m_{y}^{2}}  \\
 m_{x}  &= \frac{1}{N} \sum_{i=1}^{N} (-1)^{x_{i}}S_{i} \\
 m_{y}  &= \frac{1}{N} \sum_{i=1}^{N} (-1)^{y_{i}}S_{i}.
\end{aligned}
\label{msaf}
\end{equation}

The susceptibility corresponding to each of these order parameters is given by:

\begin{equation}
\displaystyle \chi = \frac{\langle \phi^{2} \rangle - {\langle \phi \rangle}^{2} }{k_{B}T},
\label{suscept}
\end{equation}%
where $\phi$ is any of the above order parameters and $\langle \dots \rangle$ is the thermal mean, which is the average calculated from all the meassurements taken after the steady state is reached, for a given temperature. On the other hand, the Binder cumulant is useful for estimating the critical points, and is calculated as follows:

\begin{equation}
\displaystyle g = 1- \frac{\langle \phi^{4} \rangle}{3 { \langle \phi^{2} \rangle}^{2}}
\label{binder}
\end{equation}

As is well known, the crossing point of $g_{L}-T$ curves  for different sizes $L$ locates the critical point for a second-order phase transition. For a first-order transition the cummulant presents a particular shape with a low peak at which the transition temperature can be localized. In this work we did not find evidences of a first-order behavior. Furthermore, we also used the second moment correlation length  $\xi$ to locate the critical point \cite{katzgraber2008,campbell_2019}, which is given by:

\begin{equation}
\begin{aligned}
\displaystyle  \xi  &= \frac{1}{2\sin(k_{min}/2)}  \sqrt{ \frac{\widetilde{G}(\boldsymbol{0})}{\widetilde{G}(\boldsymbol{k_{min}})} - 1 }  \\
 \widetilde{G}(\boldsymbol{k})  &= \frac{1}{N} \sum_{\boldsymbol{r}} G(\boldsymbol{r})\exp(i \boldsymbol{k.r}) ,
\end{aligned}
\label{xi1}
\end{equation}%
where $ G(\boldsymbol{r})$  the correlation function, $\boldsymbol{r} = (x,y)$ and $ \boldsymbol{k_{min}} = (2 \pi /L, 0)$. In order to associate the correlation with the leading magnetic orderings studied in this work, this formula is applied in the following form:

\begin{equation}
\begin{aligned}
\displaystyle  \xi &= \frac{1}{2\sin(\pi / L)}  \sqrt{ \frac{\langle \phi^{2} \rangle}{\langle F \rangle}-1 }, \\
F &= \frac{1}{2N} ( F_{1} + F_{2}   ),
\end{aligned}
\label{xi2}
\end{equation}%
where $F_{1}$ and $F_{2}$ are computed according to the corresponding order parameter, as follows

\begin{equation}
\begin{aligned}
\displaystyle F_{1} &= \left(\sum_{x=1}^{L}\sum_{y=1}^{L} a_{x,y} S_{x,y} \cos(2\pi x/L)\right) ^{2} + \left(\sum_{x=1}^{L}\sum_{y=1}^{L} a_{x,y}S_{x,y} \sin(2\pi x/L )\right)^{2} \\
 F_{2} &= \left(\sum_{x=1}^{L}\sum_{y=1}^{L} b_{x,y} S_{x,y} \cos(2\pi y/L)\right) ^{2} + \left(\sum_{x=1}^{L}\sum_{y=1}^{L} b_{x,y} S_{x,y} \sin(2\pi y/L)\right) ^{2}.
\end{aligned}
\label{F12}
\end{equation}%
If $\phi = m_{F}$, $\xi$ is denoted by $\xi_{F}$, and  $a_{x,y}=b_{x,y} = 1$. If $\phi = m_{AF}$, $\xi=\xi_{AF}$ and  $a_{x,y}=b_{x,y} = (-1)^{x+y}$. When  $\phi = m_{SAF}$, $\xi=\xi_{SAF}$,   $a_{x,y} = (-1)^{x}$ and $b_{x,y} = (-1)^{y}$. 

The formula of Eq.(\ref{xi2}) is valid for temperatures greater or equal to the critical temperature, where $\langle \phi \rangle = 0$. In the same way the  crossing point of curves with different sizes of the Binder Cummulant locates the critical point, different curves of $\xi/L$ are intercepted  at the transition point of a second-order phase transition (unless scaling corrections were needed). For instance, the  correlation length defined in Eq.(\ref{xi2}) depends on the lattice size $L$ at $T=T_{c}$, as follows:

\begin{equation}
\xi =  L(x^{*}+AL^{-\Delta}+\dots).
\label{xiL}
\end{equation}  

So, at $T=T_{c}$,  the limit $ \lim_{L \to \infty} \xi(L)/L$ tends to a constant, which is  $x^{*} \approx 0.905 $, for the 2D Ising model \cite{salas2000}.  On the other hand, at higher dimensions,  above the upper critical dimension, the scaling behavior is different from that of Eq.(\ref{xiL}) \cite{flores2015,young2005}. 

\section{Results}
In what follows, the value of $J_{2}$ will be expressed in units of $J_{1}$, and the critical temperature in $J_{1}/k_{B}$ units. Thus, these values are set as 1.  
\subsection{Case with ferromagnetic nearest-neighbor interactions ($J_{1} >0 $)}

For this case we considered firstly the $J_{2}$ bonds obeying the  \textbf{PDF} given in Eq (\ref{prob1}). So we have a spin lattice with ferromagnetic nearest-neighbor interactions with random bonds being ferromagnetic with probability $p$ and superantiferromagnetic with probability $1-p$.   An exploration of the order parameters $m_{F}$ and $m_{SAF}$ given in equations (\ref{mf}) and (\ref{msaf}) was done for $0 \leq p < 1$ so as to obtain the frontiers dividing phases \textbf{SAF} and \textbf{F}  with the paramagnetic one (\textbf{P}). 

\begin{figure}[htp]
\centering
\includegraphics[width=6.5cm,height=5.5cm]{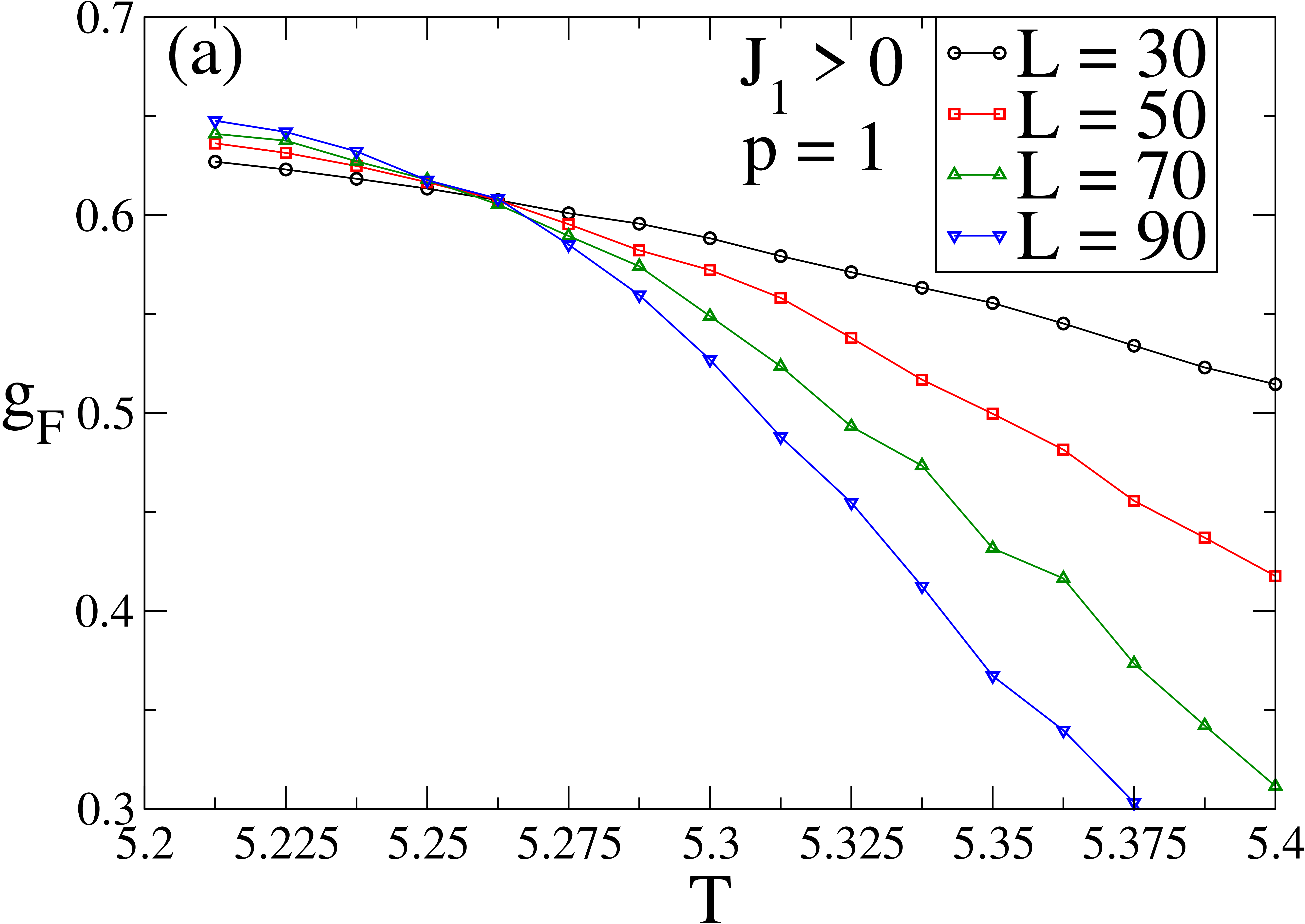}
\includegraphics[width=6.5cm,height=5.5cm]{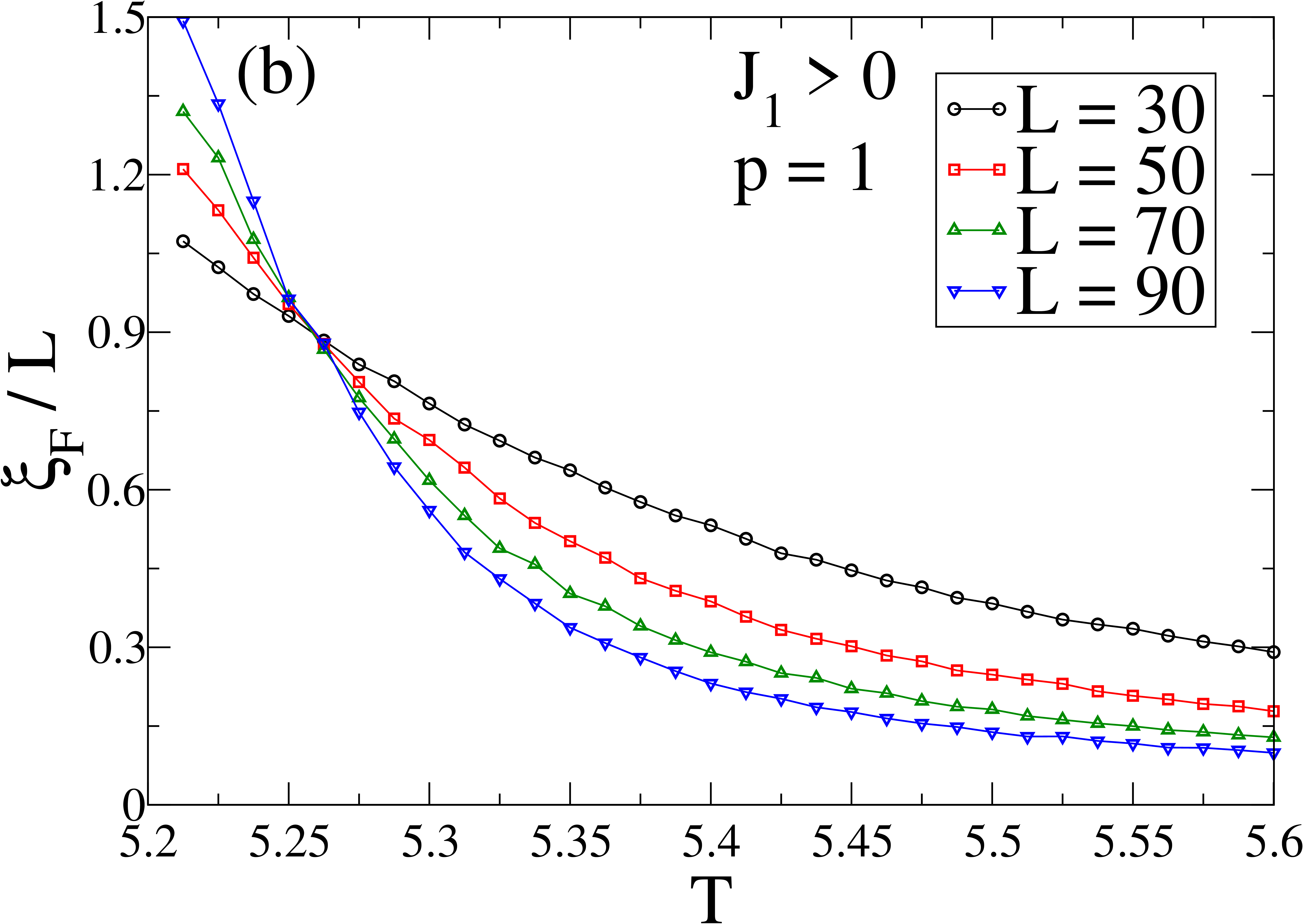}
\caption{\small{(a) Binder Cumulant; (b) and the second-order correlation length using four lattice sizes, for fixed ferromagnetic $J_{1}$ and $J_{2}$ couplings (case $p=1$ in Eq.(\ref{prob1})). The scale invariance at the critical temperature is clearly seen at the crossing point in (a) and (b). The critical temperature is $T_{c} \approx 5.26$.} } 
\label{case1_p1}
\end{figure}

\begin{figure}[htp]
\centering
\includegraphics[width=6.5cm,height=5.5cm]{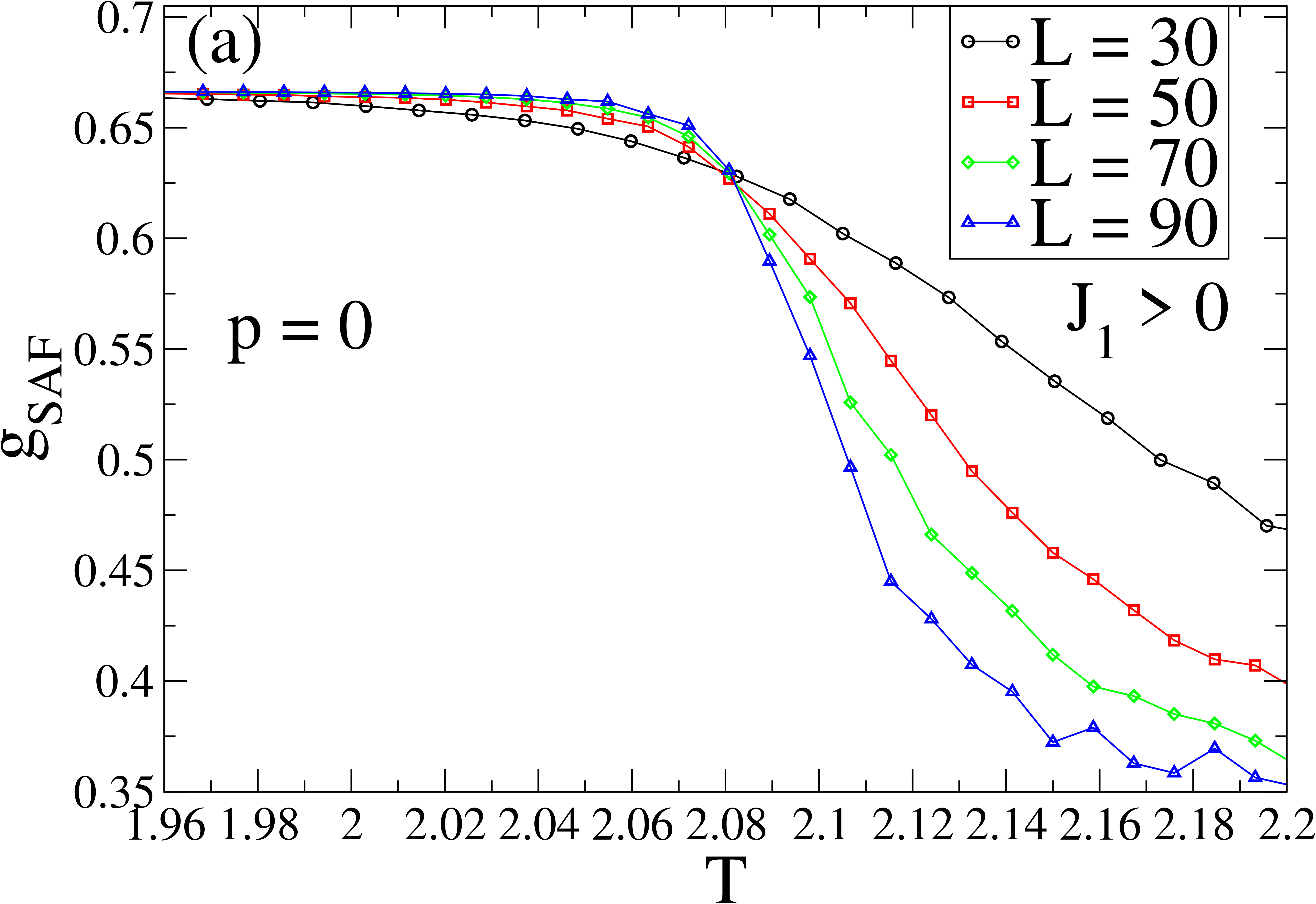}
\includegraphics[width=6.5cm,height=5.5cm]{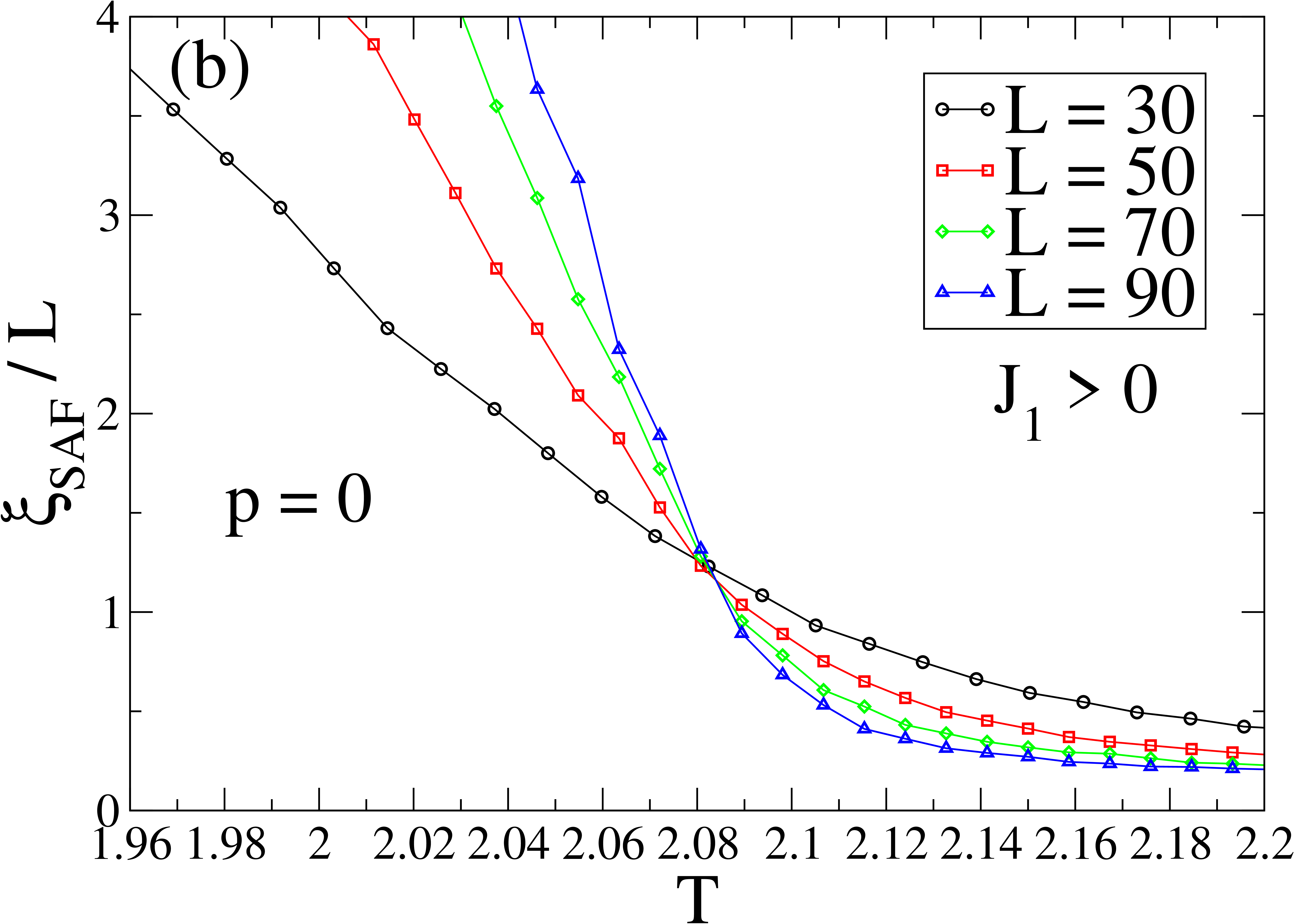}
\caption{\small{(a) Binder Cumulant; (b) and the second-order correlation length using four lattice sizes, for fixed ferromagnetic $J_{1}$ and antiferromagnetic $J_{2}$ couplings (case $p=0$ in Eq.(\ref{prob1})). Note that the critical temperature is close to $T=2.08$ (see reference \cite{malakis_2006})}.  } 
\label{case1_p0}
\end{figure}

For $p=1$, we have the trivial case of the Ising model in the square lattice with nearest- and next-nearest-neighbor interactions having a second-order phase transition at $T_{c} = 5.262 \pm 0.001$ (in agreement with reference \cite{dalton_1969}). For $p=0$, the \textbf{SAF} phase is present at low temperatures, and  the \textbf{SAF-P} transition is found  at $T = 2.083 \pm 0.002$ (the same value was obtained in \cite{malakis_2006}). Figures \ref{case1_p1} and \ref{case1_p0} show how these values were determined by locating the crossing point of respective Binder Cummulant and  correlation length. It is important to point out that the present curves of the second moment correlation length are only valid for $T \geq T_{c}$, but for the sake of visualization they are plotted for $T < T_{c}$ too. Of course, the criticality for  $p=0$ and $p=1$ is reached at equilibrium, since no competing kinetics takes place.  

\begin{figure}[htp]
\centering
\includegraphics[width=6.5cm,height=5.5cm]{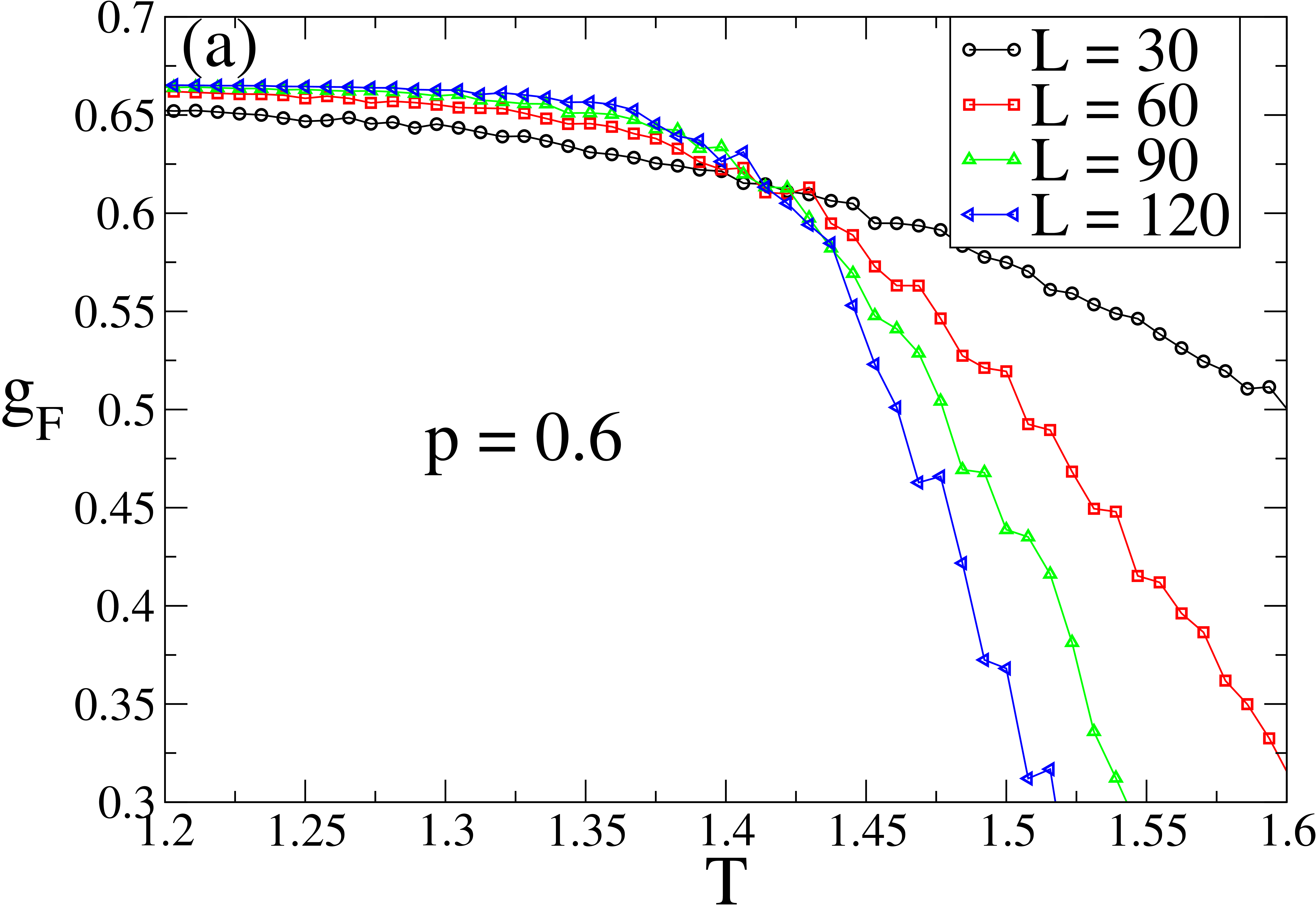}
\includegraphics[width=6.5cm,height=5.5cm]{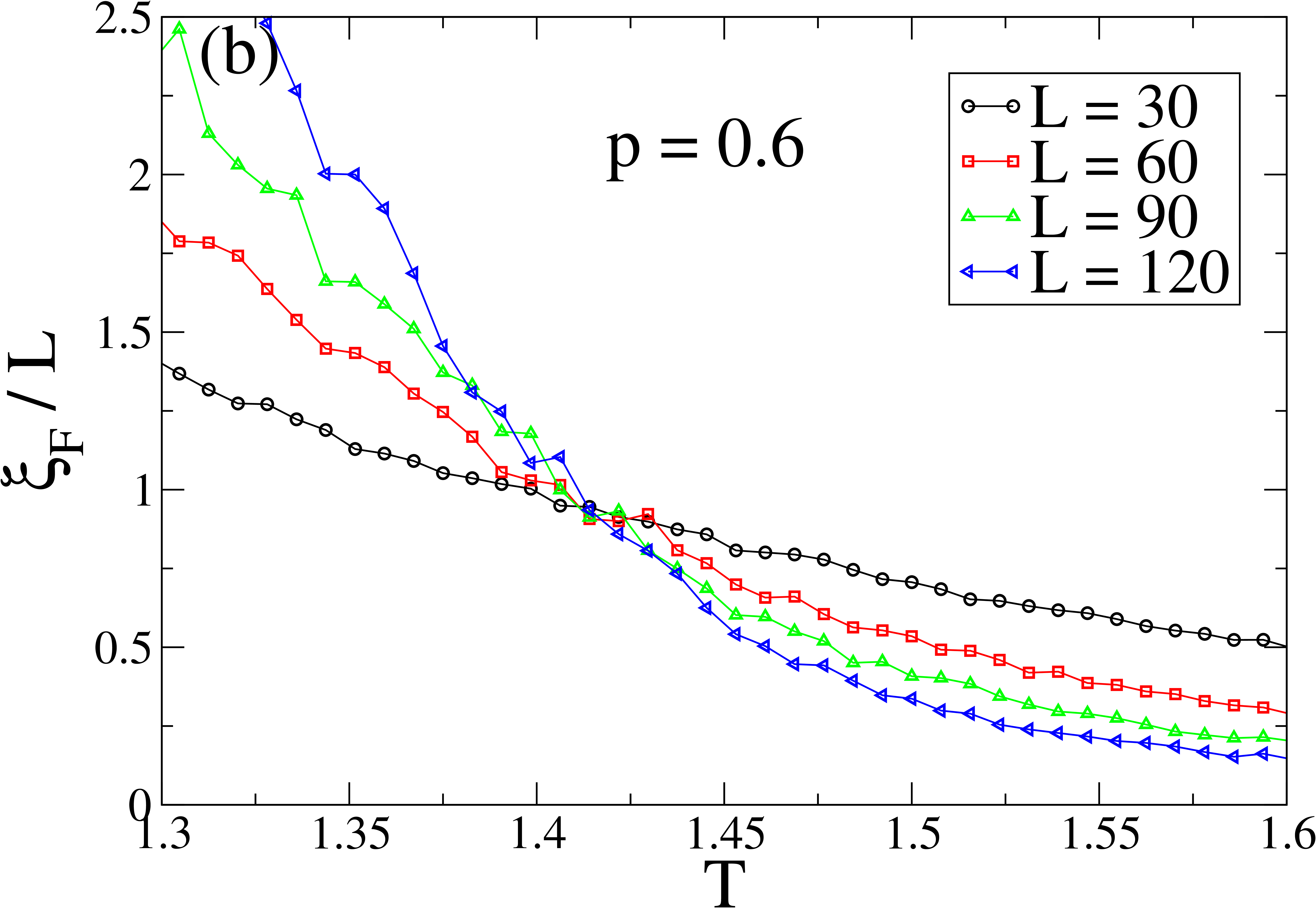}
\caption{\small{(a) Binder Cumulant; (b) and the second-order correlation length using four lattice sizes, for fixed ferromagnetic $J_{1}$ couplings, and random $J_{2}$ couplings obeying the \textbf{PDF} given in Eq.(\ref{prob1}), for $p=0.6$. This is a nonequilibrium critical steady state.  The scale invariance at the critical temperature is clearly seen at the crossing point in (a) and (b).} } 
\label{case1_p0.6}
\end{figure}

The role of $p$, for $0 < p < 1$,  is to  create a competition between \textbf{SAF} and \textbf{F}, which is stronger for its intermediate values (see  Eq.(\ref{prob1})). When $p$ is close to 1, the ferromagnetic order is present as Fig.\ref{case1_p0.6} shows. There it is observed the Binder Cummulant $g_{F}$ and the second moment correlation length divided by the lattice size  $\xi_{F}/L$ as functions of the temperature, corresponding to the ferromagnetic order parameter $m_{F}$, for $p=0.6$. Four curves, for sizes $L=30,60,90,120$, colapse  at a  point in Fig.\ref{case1_p0.6}(a) and (b) locating  the critical temperature. Both figures agree with the value $T_{c} = 1.410 \pm 0.005$.  This is a clear signal of second-order phase transition. However, the scale invariance at the critical point seems to be better visualized using the correlation length. 

\begin{figure}[htp]
\centering
\includegraphics[width=6.5cm,height=5.5cm]{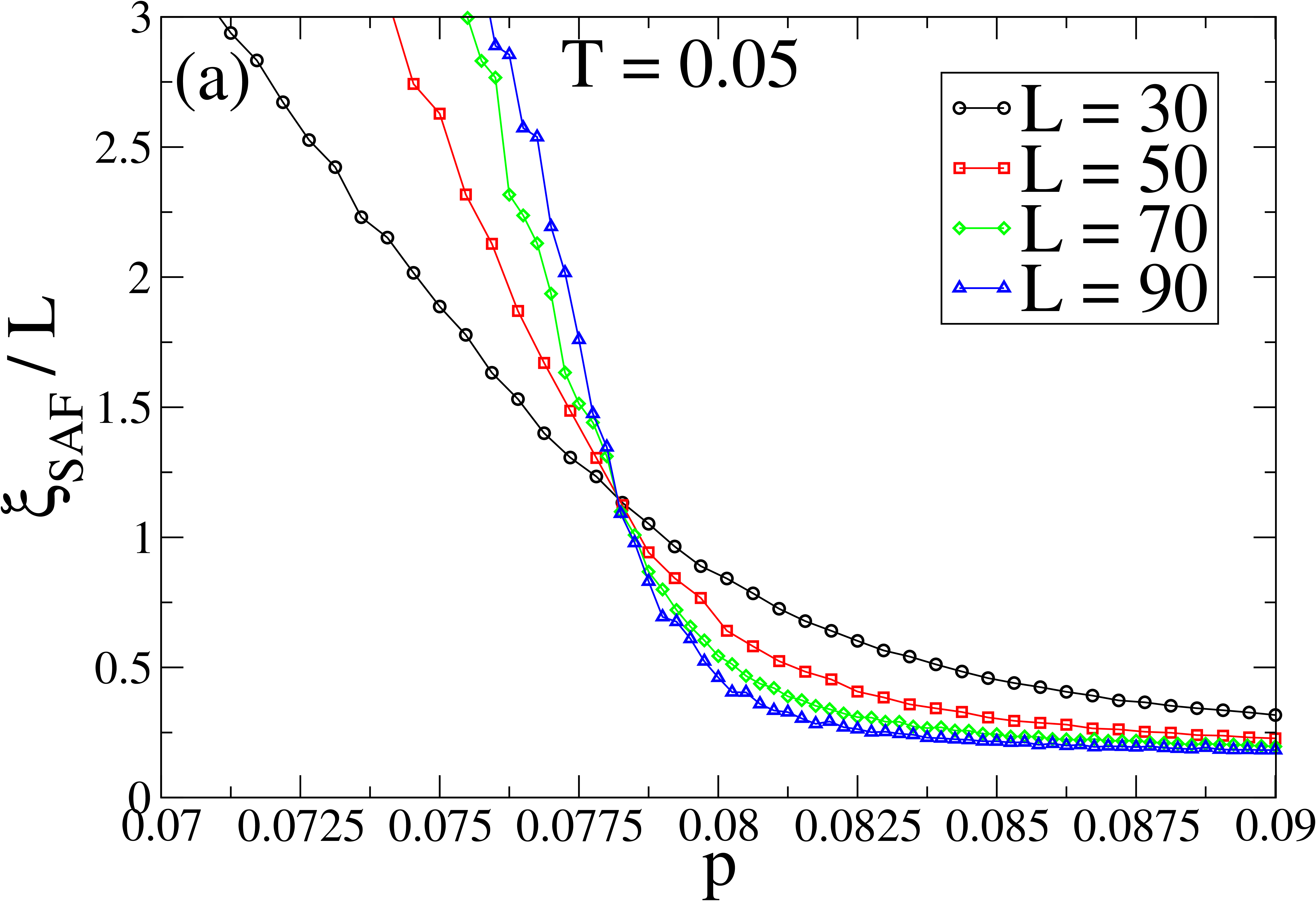}
\includegraphics[width=6.5cm,height=5.5cm]{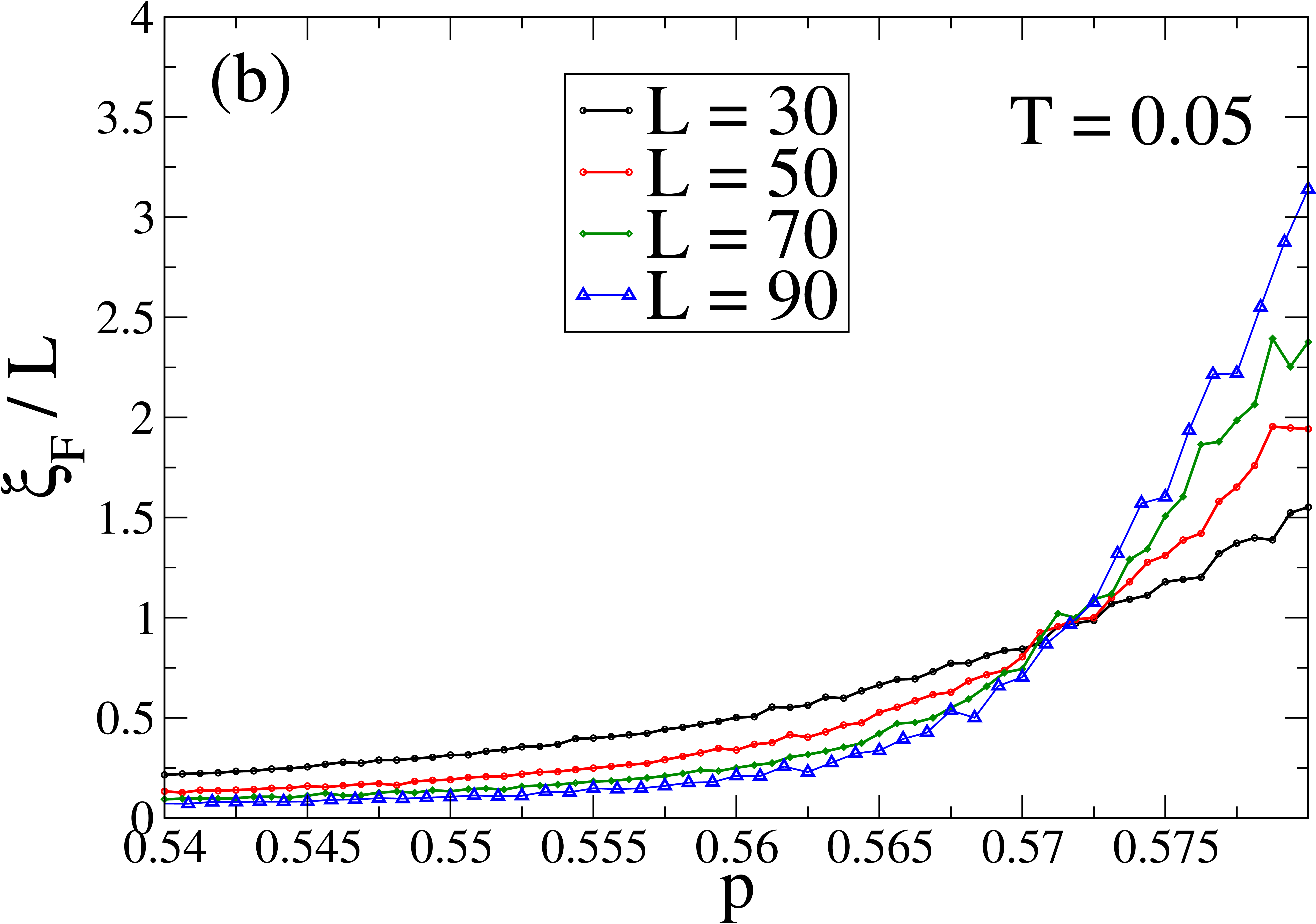}
\caption{\small{Correlation length scanned at a fixed very low temperature,  for fixed ferromagnetic $J_{1}$ couplings, and random $J_{2}$ couplings obeying the \textbf{PDF} given in Eq.(\ref{prob1}). In (a) the critical probability of the \textbf{SAF-P} transition is about  $p=0.078$. In (b) the critical probability of the \textbf{F-P} transition is about $p=0.57$.  } } 
\label{case1_T0.05}
\end{figure}

\begin{figure}[htp]
\centering
\includegraphics[width=9cm,height=6cm]{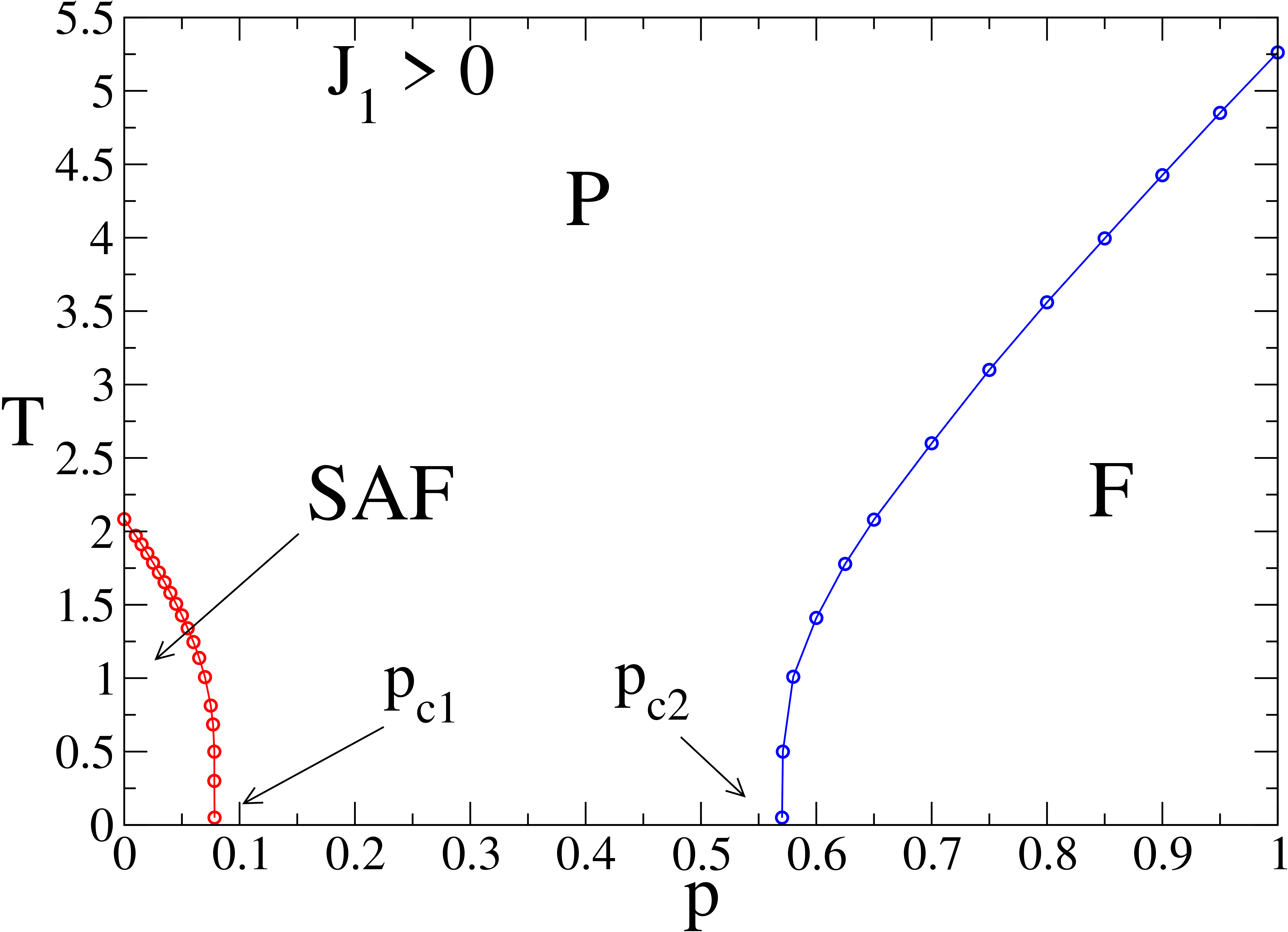}
\caption{\small{Phase diagram of the model given by the Hamiltonian given in Eq.(\ref{ham}),   for fixed ferromagnetic $J_{1}$ couplings, and random $J_{2}$ bonds obeying the \textbf{PDF} given in Eq.(\ref{prob1}). Critical points for $p=0$ and $p=1$ are at equilibrium, whereas for $ 0 < p < 1$, the frontier points are at nonequilibrium steady states.  } } 
\label{case1_df1}
\end{figure}

In Fig.\ref{case1_T0.05} is shown  the correlation length $\xi_{SAF}$ and $\xi_{F}$  scanned through the probability $p$, for a very low temperature $T = 0.05$. In Fig.\ref{case1_T0.05}a the crossing point is located at $p_{c} = 0.0783 \pm 0.0004$, whereas in Fig.\ref{case1_T0.05}b we have $p_{c}= 0.571 \pm 0.006$. Consequently, there is a paramagnetic gap between the \textbf{SAF-P} and \textbf{F-P} second-order frontiers.  In Fig.\ref{case1_df1} the resulting phase diagram is exhibited, where all points were obtained using the correlation length and the Binder Cummulant. Note the paramagnetic gap for $  p_{c1} \leq p  \leq p_{c2} $, where $p_{c1} \approx 0.078$ and $p_{c2} \approx 0.57$.   

If $J_{2}$ obeys the \textbf{PDF} given in Eq.(\ref{prob2}) (a double Gaussian),  any finite value of the standard deviation $\sigma$ destroys the \textbf{SAF} phase. In what follows, $\sigma$ will be expressed in units of $J_{1}$. Note that the \textbf{PDF} given in Eq.(\ref{prob1}) is the limit of the double Gaussian given in Eq.(\ref{prob1}), when $\sigma \to 0$. Interestingly, the effect of a very small $\sigma$ is to shift the extreme critical points of the phase diagram from $p_{c} \approx  0.57$ to  $p_{c} \approx 1/3$ and from $T_{c}(p=1) \approx 5.26$ to $T_{c}(p=1) \approx 3.57$. We confirmed that this shift exists even for a very small  value of $\sigma$, such as $\sigma = 0.001$. 

In Fig.\ref{case1_df1} the phase diagram is presented for $\sigma = 0.1$. There, the extreme points of the frontier are $p_{c} = 0.3310 \pm 0.0005$,  for $T=0.005$, and for $p=1$, $T_{c} = 3.5730 \pm 0.0004$. To understand the effect of the standard deviation  on $p_{c}$ we show in Fig\ref{pc1}a  how $p_{c}$ evolves with $\sigma$. It remains constant for $0 < \sigma < 0.3$, because the gaussians of the \textbf{PDF} do not overlap, as Fig\ref{pc1}b shows. Then, it increases from $\sigma \approx 0.3$, since the gaussians of the \textbf{PDF} increases their overlap. For  $\sigma = 2.5$, $p_{c}  =0.987$(see Fig\ref{pc1}b). Accordingly, for $\sigma \gtrsim 2.5$, $p_{c}$ reaches its maximum value, which is one. So, the ferromagnetic area of the phase diagram is totally reduced, since  the randomness caused by $\sigma$ is so strong, such  that any amount of  the second Gaussian in Eq.(\ref{prob2}) (any amount of ($1-p$)), would destroy the ferromagnetic phase (see Fig\ref{pc1}).  On the other hand, the effect of $\sigma$ on $T_{c}(p=1)$ is presented in Fig.\ref{tc1}. 

In Fig.\ref{tc1}a it can be observed the discontinuous fall in the critical temperature $T_{c}(p=1)$ when $\sigma$ assumes infinitesimal values, remaining constant for $ 0 < \sigma < 0.5$. Then it increases slightly before reaching its maximum for $\sigma \approx 1$. Then $T_{c}(p=1)$ decreases. The extension of the Gaussian distribution for these three different stages  is exhibited in Fig.\ref{tc1}b, for three representative values of $\sigma$. Note that the critical temperature is unaltered when the probability for negative values of $J_{2}$ is zero. However, if the left tail of the Gaussian includes the interval $ -2 \lesssim J_{2} < 0$, $T_{c}(p=1)$ increases. If this  interval of negative $J_{2}$ values  is extended, the effect of $\sigma$ is to decrease the critical temperature. In Fig.\ref{tc1}a  the critical temperature is plotted  up to  $\sigma = 2.5$, when $p_{c}$ assumes the value 1.

\subsection{Case where $J_{1}<0$}

Now we treat the case  $J_{1} < 0$, where the nearest-neighbor interactions are antiferromagnetic. If the \textbf{PDF} given in Eq.(\ref{prob3}) is obeyed by the $J_{2}$ bonds, these  are antiferromagnetic  with probability $1-p$, and  zero with probability $p$. So,  $p$ measures the degree of dilution of $J_{2}$ bonds. This particular case was studied in equilibrium with quenched randomness in reference \cite{yining_2018}, where $x$ stood for our probability of dilution $p$, as mentioned before in the Introduction. 

The phase diagram  we obtained is shown in Fig.\ref{case1_df3}, where the \textbf{SAF-P} and \textbf{AF-P} frontiers are separated by paramagnetic gap for $  p_{c1} \leq p  \leq p_{c2} $, where $p_{c1} \approx 0.309$ and $p_{c2} \approx 0.69$. The critical temperatures at the extremes of the phase diagram are $T_{c}(p=0) \approx 2.08$ (in agreement with \cite{malakis_2006}), and  $T_{c}(p=1) = 2.269... $ (the exact 2D Ising one).

Phase diagram in Fig.\ref{case1_df3} can be compared with that exhibited in Fig.5a of reference \cite{yining_2018}. There, the Neel phase is our \textbf{AF} phase and the strip antiferromagnetic one is our \textbf{SAF} phase. The topology is quite similar. Nevertheless, we did not find a spin-glass phase in the gap area close to $T=0$, as in reference \cite{yining_2018}. At least, not with the spin-glass order parameter averaging the correlation between the spins of two copies of the system (see Eq.(6) in \cite{yining_2018}). 

Furthermore, the frontiers we obtained exhibit humps, which do not appear with quenched randomness in that paper. In the present case, the \textbf{SAF-P} frontier reaches its maximum at $(p,T_{c}) \approx (0.075,2.196)$, and at $(p,T_{c}) \approx (0.925,2.364)$, for the \textbf{AF-P} frontier, as shown in Fig.\ref{case1_p0075}. These humps may be caused  by the particular type of competing kinetic used in this work, which  causes an increase of the critical temperature for values of $p$ close to zero and $1$.

Finally, we applied the \textbf{PDF} given in Eq.(\ref{prob4}) for the $J_{2}$ bonds, with $J_{1} <0$. In this case $J_{2}$ is distributed by two gaussians centered at $J_{2}=0$ and  $J_{2}=J_{1}$, with probability $p$ and $1-p$, respectively. When $\sigma \to 0$ the  bimodal given in Eq.(\ref{prob3}) is recovered. In Fig\ref{case1_df4} is presented the phase diagram for $\sigma = 0.1$. Again we observe that the \textbf{SAF} is not present. This happens even when $\sigma$
is infinitesimal. Thus, only the \textbf{AF} survives in an area enclosed by a second-order frontier, where $p_{c} \approx 0.692$ and $T_{c}(p=1)$ is the 2D Ising critical temperature. The evolution of $p_{c}$ with $\sigma$ is shown in Fig.\ref{pcc1}a, where we may see how $p_{c}$ falls from $p_{c} \approx 0.692$  down to $p_{c} \approx 0.383$, when $\sigma$ assumes an infinitesimal value. Then, $\sigma$ does not affect the value of $p_{c}$ until  $\sigma \gtrsim 0.35$.

The \textbf{AF} area disappears when $p_{c} = 1$, this happens for $\sigma \gtrsim 1.3$. In   Fig.\ref{pcc1}b is shown that $p_{c}$ is not affected by  $\sigma$ when the tails of the two gaussians die for $-1 < J_{2} < 1$. For  $\sigma \lesssim 0.35$, the tails of the two gaussians start to include $|J_{2}| > 1$, so $p_{c}$ begins to increase. Similarly, in Fig.\ref{tcc1} is shown the evolution of $T_{c}(p=1)$ with $\sigma$, until $p_{c}$ reaches the value 1. It confirms that the frontier  exhibited in Fig.\ref{case1_df4} is unaltered when $\sigma < 0.35 $.

\subsection{The amplitude $x^{*}_{SAF}$}

The amplitude  $x^{*}_{SAF}$ in Eq.(\ref{xiL})  has attracted our attention. We may observe that the  crossing point of $\xi_{SAF}/L$ curves gives different values of $x^{*}_{SAF}$ when $J_{2}=-J_{1} <0$ and $J_{2} = J_{1} < 0$. This is exhibited in Fig.\ref{curves_p0J1J2}, where  the point of intersection provides $T_{c}(p=0)$ in Figures \ref{case1_df1} and \ref{case1_df3}.    In Fig.\ref{curves_p0J1J2}a the uncertainty of the crossing point also gives $x^{*}_{SAF}=1.210 \pm 0.005$, and in Fig.\ref{curves_p0J1J2}b $x^{*}_{SAF} = 1.483 \pm 0.008$. However, the critical temperatures are just the same $T_{c}(p=0) \approx 2.082$, whithin the error. 

So, though both critical points are just at the same \textbf{SAF-P} critical temperature, the way $J_{1}$ and $J_{2}$ bonds are configured to form the \textbf{SAF} phase gives a slight different value of $x^{*}_{SAF}$. To reinforce this idea, we analysed the effect of the bimodal distributions given in Eqs.(\ref{prob1}) and (\ref{prob3}) on this superanfiterromagnetic amplitude. For the case $J_{1} > 0$, the \textbf{SAF} phase is destroyed for a small value of $p$ (see Fig. \ref{case1_df1}), so no relevant change in $x^{*}_{SAF}$ is observed. However, when $J_{1} < 0 $, the degree of dilution of $J_{2}$ changes this amplitude, as suggested by an \textit{ad-hoc} finite-size criterium. 

To show it, we have determined the interval of uncertainty of $x^{*}_{SAF}$ in Fig.\ref{extrap1}, for $p=0$, using  Eq.(\ref{xiL}). We used small sizes of $L$, such as $L = 8,10,12,14,16,18,20,24,28$, for which the second term in  Eq.(\ref{xiL}) is relevant.  The limits of  $x^{*}_{SAF}$ were determined by extrapolating to $L \to \infty$ the line of points of 
$\xi_{SAF}/L$ versus $L^{-\Delta}$, each one corresponding to the limits of the critical temperature found in the crossing point of Fig.\ref{curves_p0J1J2}b. 

The values of $\Delta$ were scanned in order  to  get the adequate ones to adjust the points in a straight line. So, the two extrapolated lines in Fig.\ref{extrap1} intersect to the vertical axis at points which constitute  the limits of the uncertainty of $x^{*}_{SAF}$.  Accordingly, we found  $1.440 < x^{*}_{SAF} < 1.497 $. A similar extrapolation was performed for the transition point at $T=0.4$ of the left frontier in Fig.\ref{case1_df3}. There we have $ 0.30882 < p_{c} < 0.30898 $, according to the uncertainty of the crossing point shown in Fig. \ref{curves_T04}a. 

The resulting interval of the amplitude is shown in Fig.\ref{curves_T04}b. Thus, $1.53 < x^{*}_{SAF} < 1.55 $, which does not intersect  the interval found for $p=0$. Therefore, this finite-size analysis suggest that  $x^{*}_{SAF}$ also changes with the dilution of $J_{2}$. Of course, more rigorous numerical treatments need to be done to confirm this affirmation. 

\section{Conclusions}

We have studied the steady states of a nonequilibrium $J_{1}-J_{2}$ Ising model with random $J_{2}$ bonds obeying bimodal and double Gaussian probability distribution functions, for ferromagnetic and antiferromagnetic $J_{1}$ couplings, in a competing kinetics. 

All the phase diagrams presented second-order frontiers dividing the paramagnetic phase (\textbf{P}) with some of the following orders : the \textbf{F}, \textbf{AF} or \textbf{SAF} order. Other authors found similar topologies for these phase diagrams obtained at equilibrium using quenched randomness with the bimodal distributions considered in this work.  We observed that the  \textbf{SAF} phase can not exist when $J_{2}$ obeys a continuous probability distribution. 

Also, no universality breaking was observed for the critical exponents of the corresponding order parameters  and their  susceptibilities. However, we found that the amplitude $x^{*}_{SAF}$ of the correlation length at the   \textbf{SAF-P} transition  is not only different to that of 2D pure Ising model, but may change if the degree of dilution of $J_{2}$ bonds is relevant. This was shown by calculating  the uncertainties of $x^{*}_{SAF}$. As far as we know, no other work has calculated this particular amplitude. It would be interesting to perform the same calculations with quenched randomness in order to compare  the obtained $x^{*}_{SAF}$ values. This will be done in a future work. 

\vspace{1.0cm}
\textbf{ACKNOWLEDGEMENTS}

 This work was partially supported by CNPq - Grants: 408787/2018-0 and 306569/2018-3 (Brazilian Research Agency). We also thank  Ivan Scivetti for fruitful discussion.

\vskip \baselineskip
\vskip \baselineskip

\vskip \baselineskip
\vskip \baselineskip

\vskip \baselineskip
\vskip \baselineskip

\vskip \baselineskip
\vskip \baselineskip

\vskip \baselineskip
\vskip \baselineskip

\vskip \baselineskip
\begin{figure}[htp]
\centering
\includegraphics[width=9cm,height=6cm]{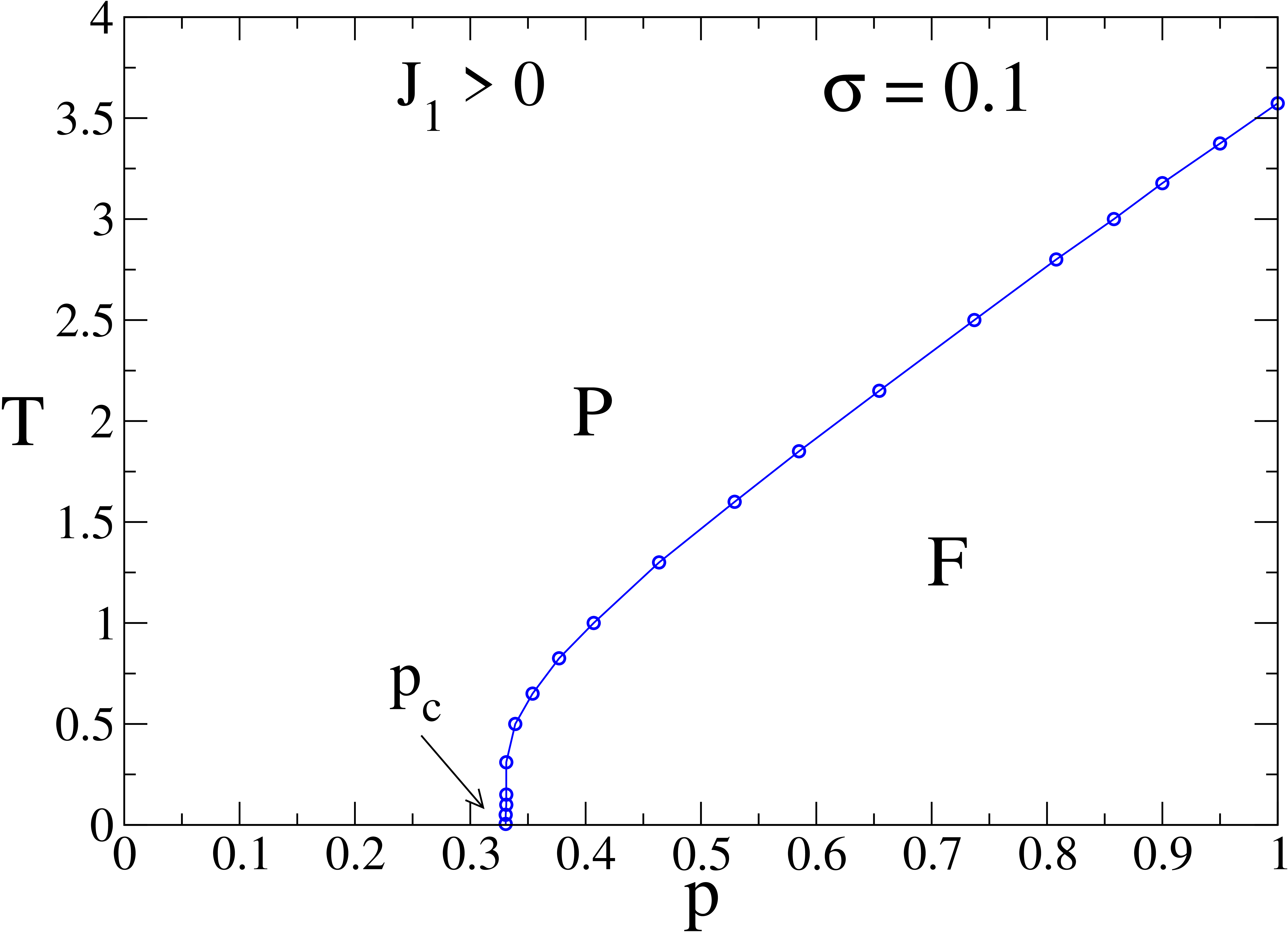}
\caption{\small{Phase diagram of the model given by the Hamiltonian given in Eq.(\ref{ham}),   for fixed ferromagnetic $J_{1}$ couplings, and random $J_{2}$ bonds obeying the \textbf{PDF} given in Eq.(\ref{prob3}), when $\sigma = 0.1$. The vertical axis is the temperature and the horizontal one is the probability $p$, which is the weight of the first gaussian  in Eq.(\ref{prob2}). There is only  one frontier separating the \textbf{F} and \textbf{P} phases, since any amount of  $\sigma$ causes de destruction of the \textbf{SAF} phase.   } } 
\label{case1_df2}
\end{figure}
\vskip \baselineskip

\vskip \baselineskip

\begin{figure}[htp]
\includegraphics[width=7cm,height=5.5cm]{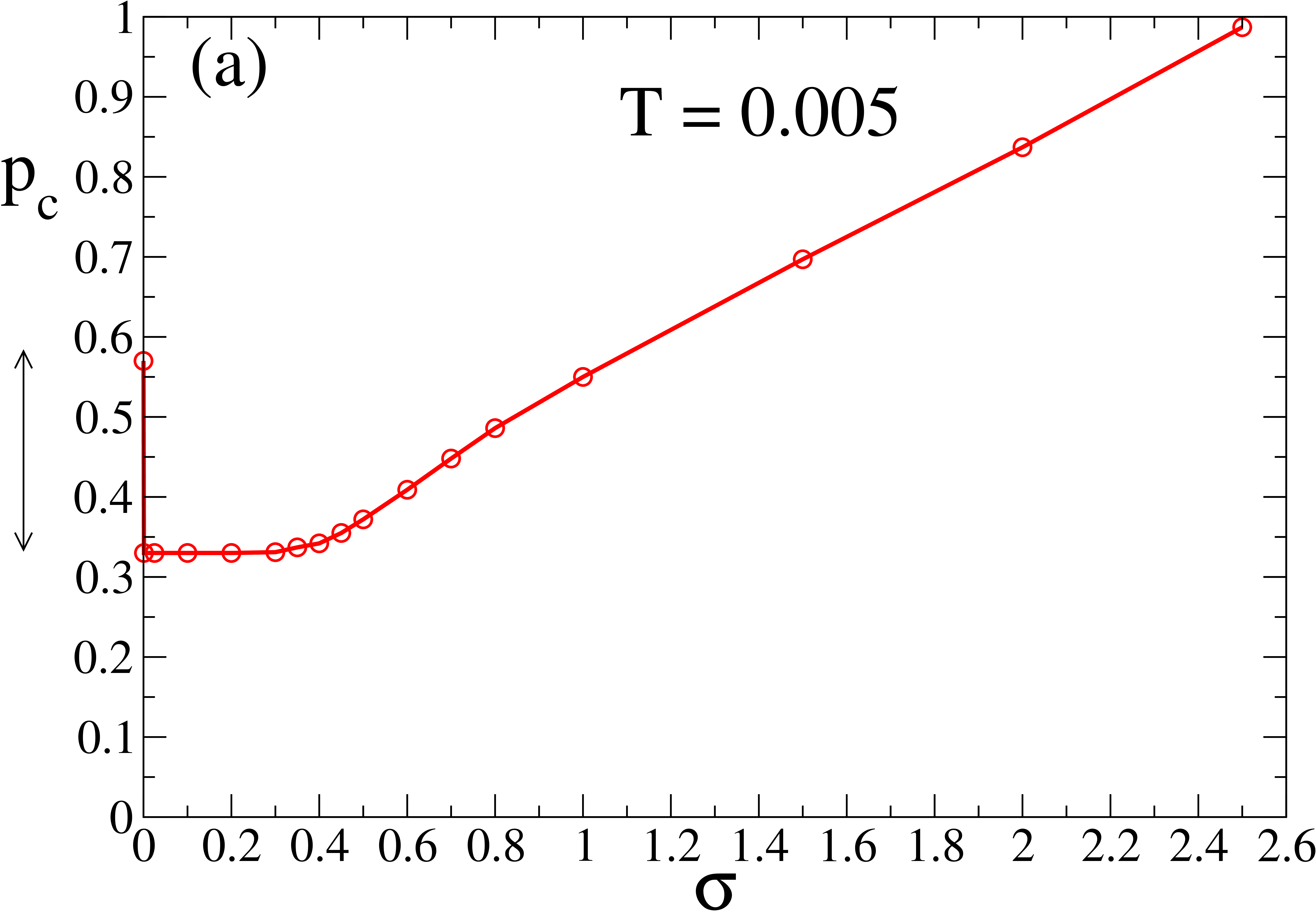}
\includegraphics[width=7cm,height=5.5cm]{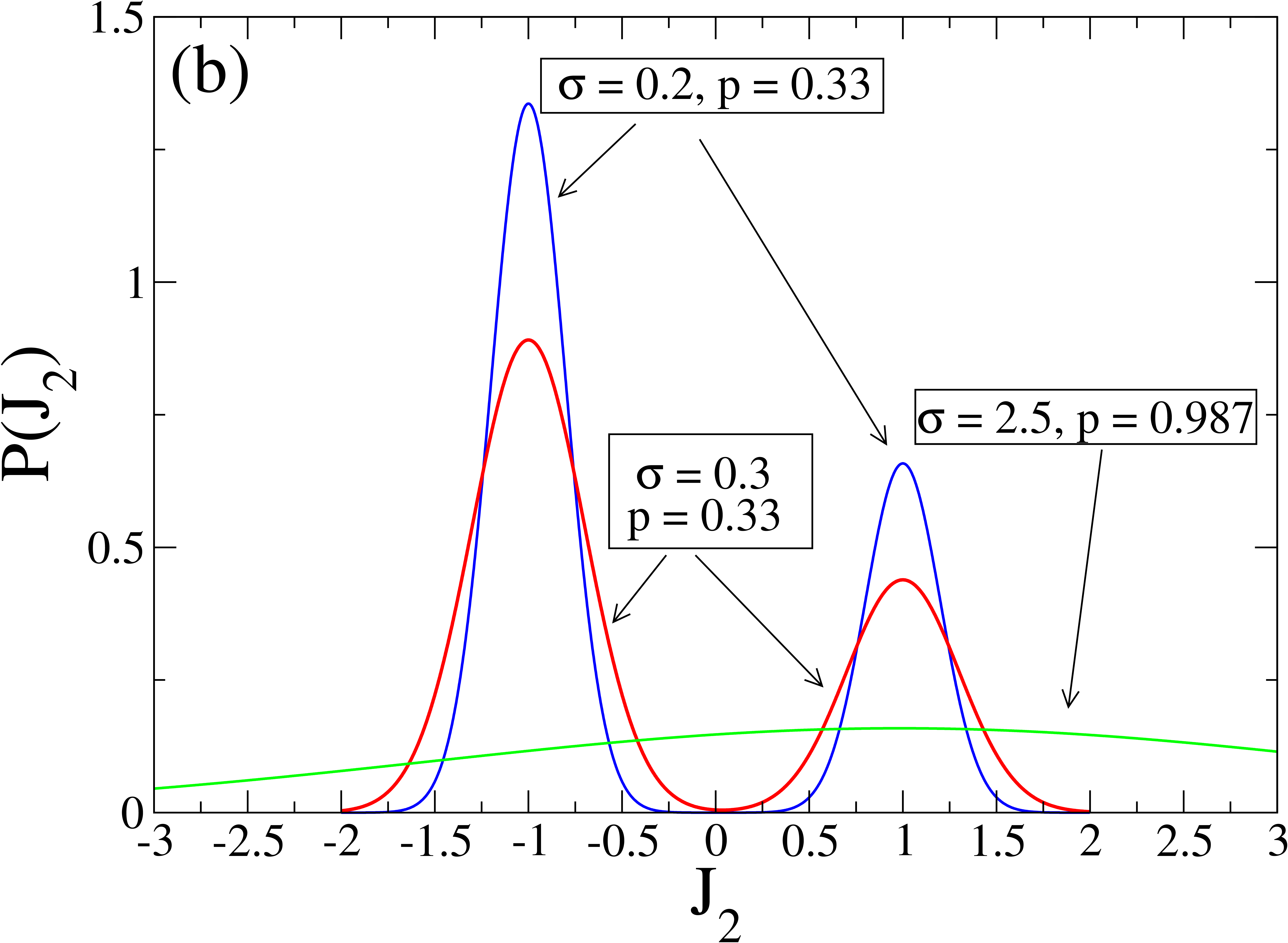}
\caption{\small{ (a) The critical probability of the phase diagram shown in Fig\ref{case1_df2} versus the parameter $\sigma$ given in Eq.(\ref{prob2}). Note the discontinuous fall in $p_{c}$ when $\sigma$ assumes a finite value.  In (b) is shown the \textbf{PDF} given in Eq.(\ref{prob2}) for three different values of the pair ($p$, $\sigma$). Note that there is a weak overlap of the two peaks, for $p_{c}=0.33$ and $\sigma = 0.3$  causing the first inflection point in the curve shown in (a). For $p_{c}=0.987$ and $\sigma = 2.5$ the left  gaussian  is just irrelevant, and  the right one is very extended.   } } 
\label{pc1}
\end{figure}
\vskip \baselineskip
\break
\vskip \baselineskip

\begin{figure}[htp]
\includegraphics[width=7cm,height=5.5cm]{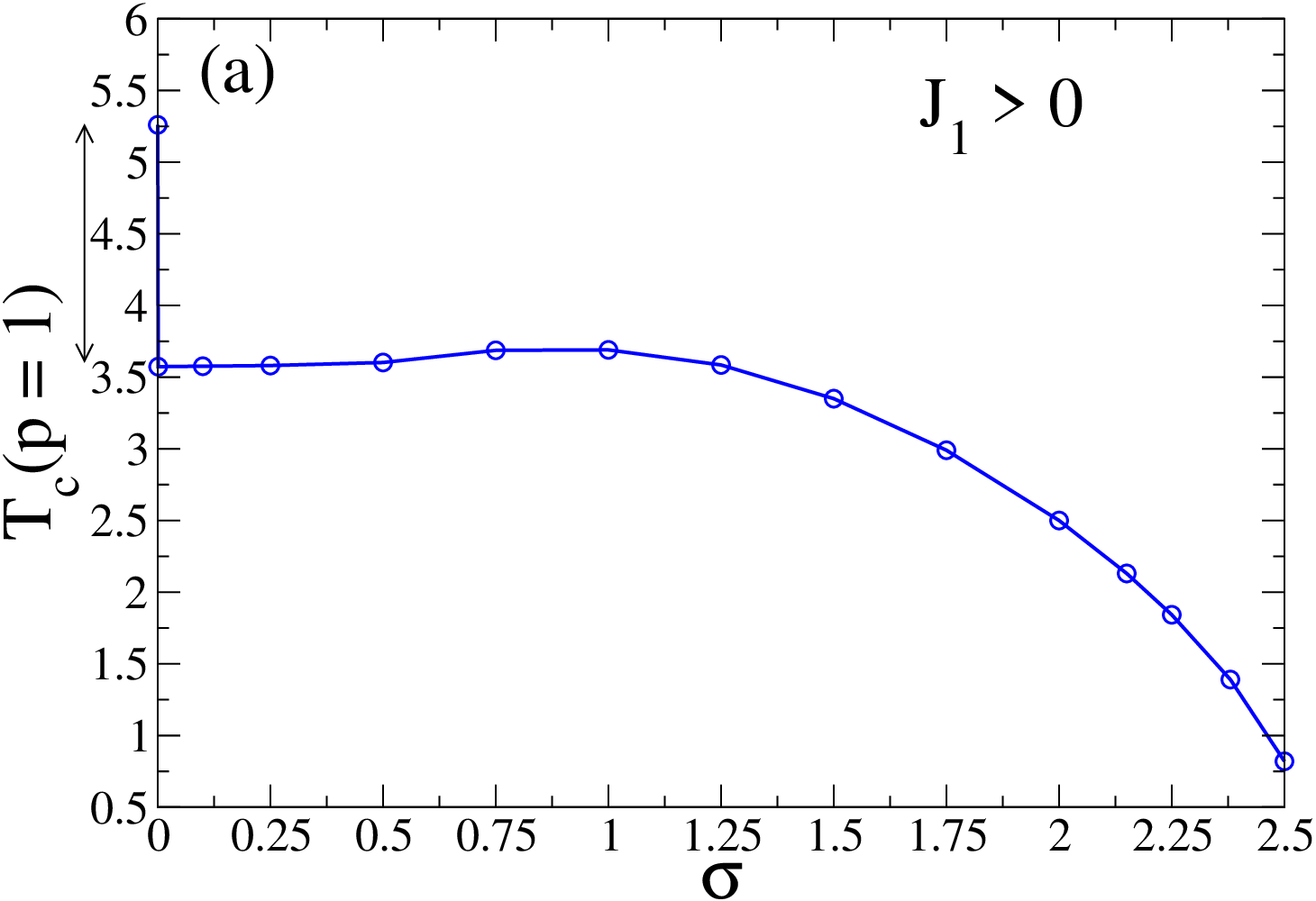}
\includegraphics[width=7cm,height=5.5cm]{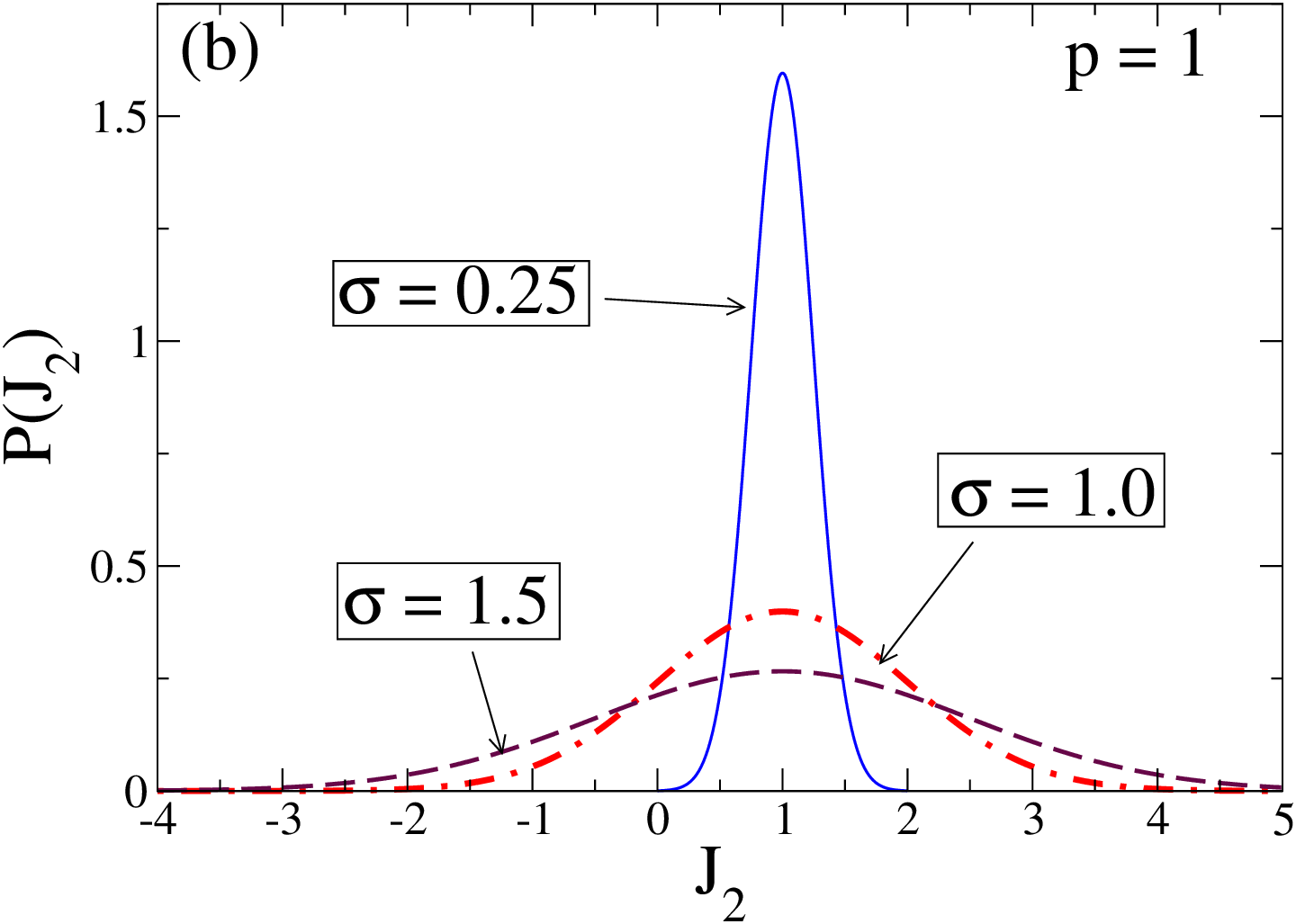}
\caption{\small{ (a) The critical temperature of the phase diagram shown in Fig\ref{case1_df2} versus the parameter $\sigma$ given in Eq.(\ref{prob2}), for $p=1$. Note the discontinuous fall in $T_{c}$ when $\sigma$ assumes a finite value.  In (b) is shown the \textbf{PDF} given in Eq.(\ref{prob2}) for three different values of $\sigma$, for $p=1$. Note that the critical temperature reaches its maximum for $\sigma = 1$, and decreases when the left tail of the gaussian include more negative values of $J_{2}$.   } } 
\label{tc1}
\end{figure}
\vskip \baselineskip

\vskip \baselineskip
\begin{figure}[htp]
\centering
\includegraphics[width=9cm,height=6cm]{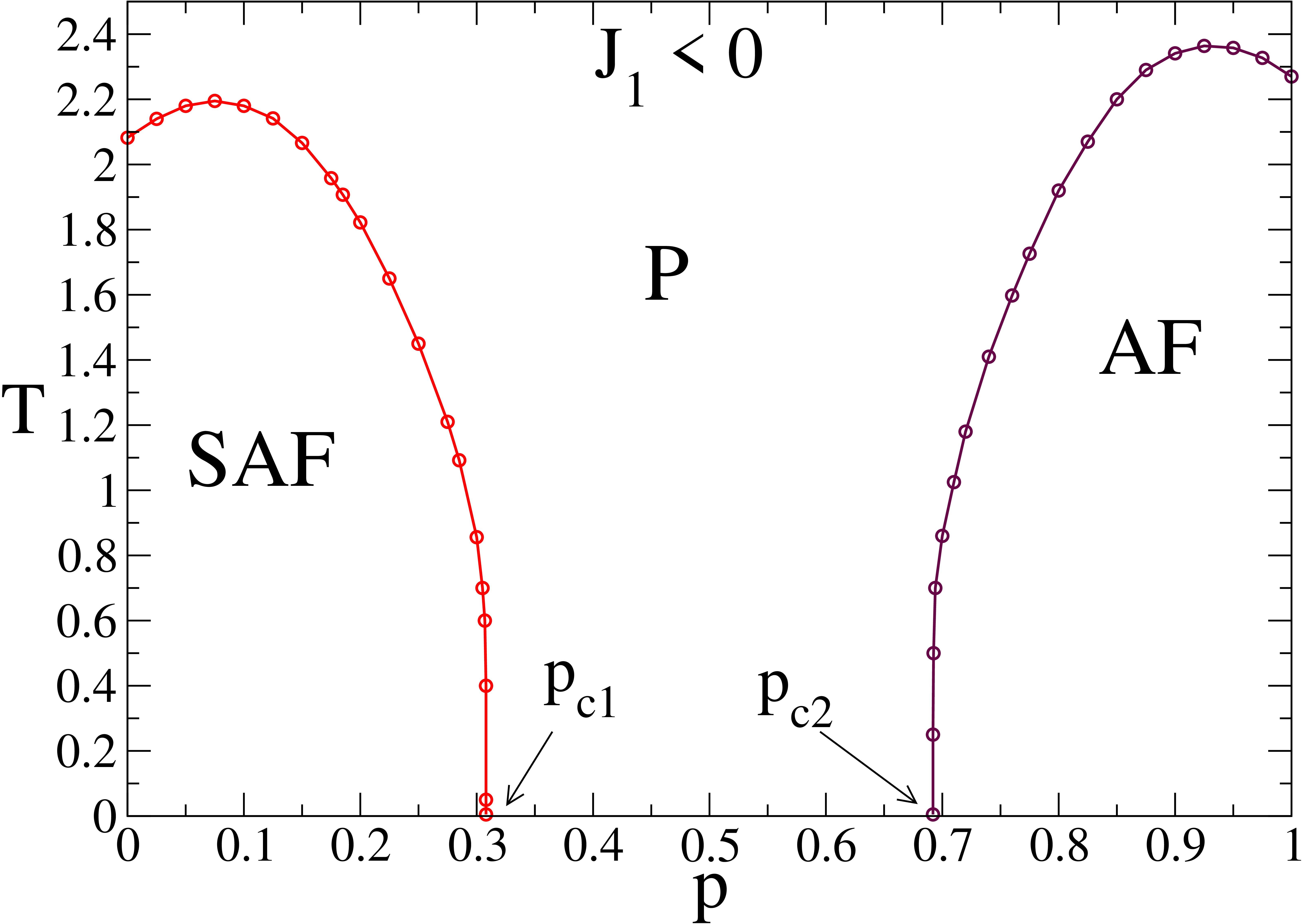}
\caption{\small{Phase diagram of the model given by the Hamiltonian given in Eq.(\ref{ham}),   for fixed antiferromagnetic $J_{1}$ interactions, and random $J_{2}$ bonds obeying the \textbf{PDF} given in Eq.(\ref{prob3}). The maxima at both frontiers are not present in the equivalent phase diagram of Fig.5a of reference \cite{yining_2018}, which was obtained with quenched randomness at equilibrium.   } } 
\label{case1_df3}
\end{figure}
\vskip \baselineskip

\vskip \baselineskip
\begin{figure}[htp]
\centering
\includegraphics[width=6.5cm,height=5.5cm]{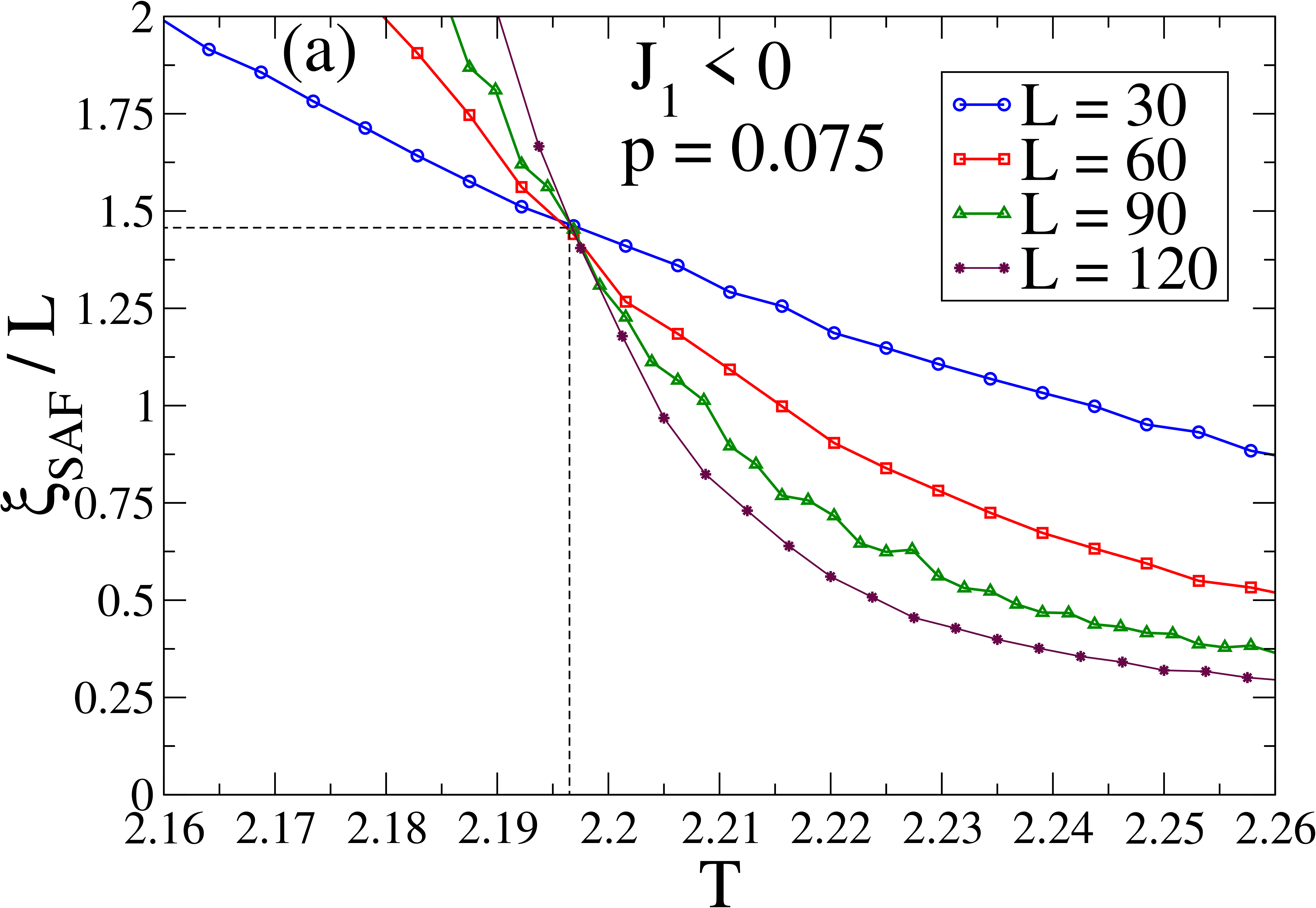}
\includegraphics[width=6.5cm,height=5.5cm]{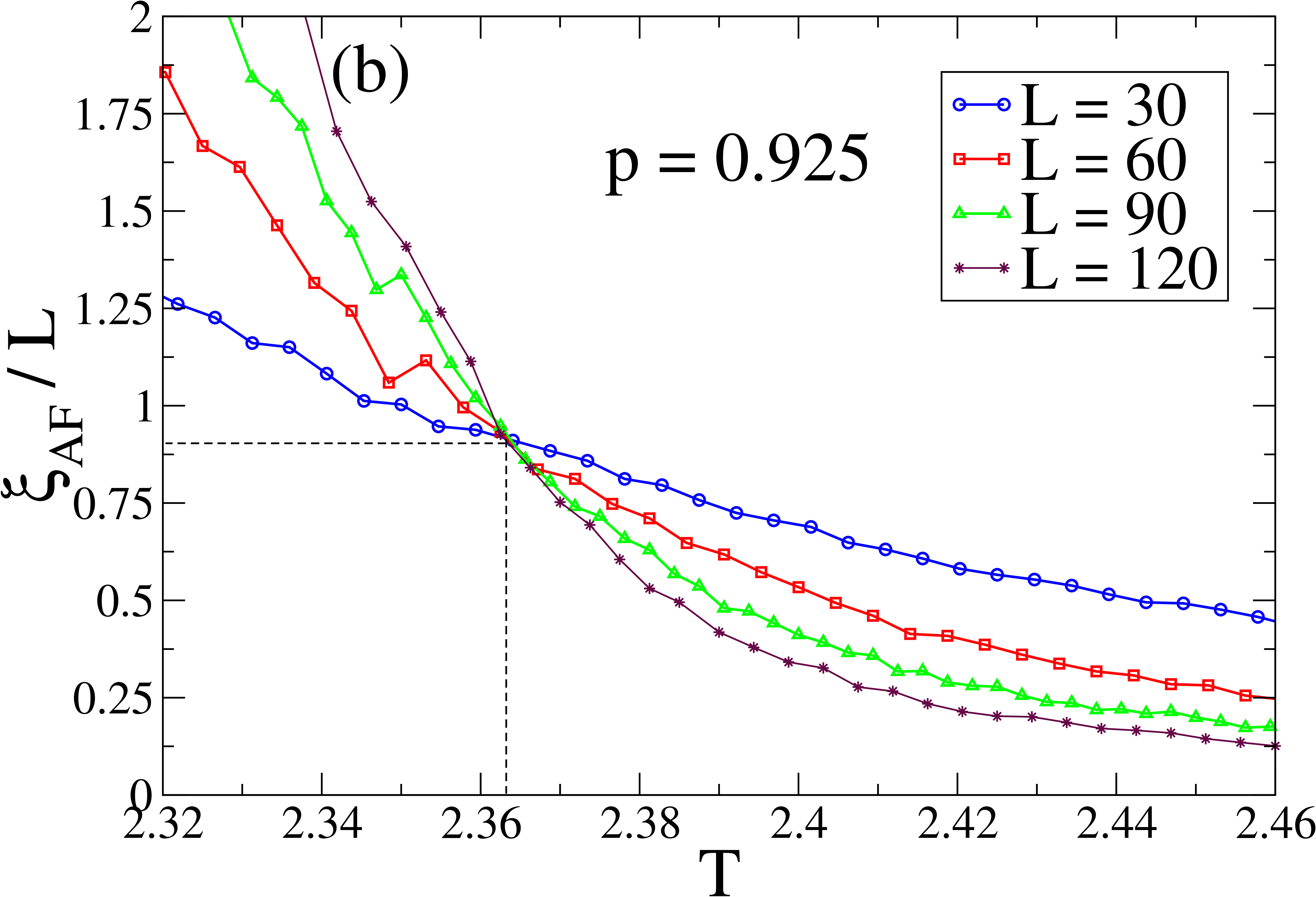}
\caption{\small{(a) Curves of $\xi_{SAF}/L$, where the crossing point is at $T= 2.196 \pm 0.003$ (above $T_{c}(p=0)$); (b) Curves of $\xi_{AF}/L$, where the croosing point is at $T = 2.364 \pm 0.004 $ (above $T_{c}(p=1)$). These two crossing points correspond to the maximum of each frontier in Fig.\ref{case1_df3}. Furthermore, in (a)  $\xi_{SAF}/L$ is   $1.467 \pm 0.008$ at the point of intersection , whereas in (b)  $\xi_{AF}/L$ is $0.904 \pm 0.003$ (which is the Ising universal amplitude $x^{*}$ \cite{salas2000}). Dashed lines are guides to the eye.   } } 
\label{case1_p0075}
\end{figure}
\vskip \baselineskip

\vskip \baselineskip
\begin{figure}[htp]
\centering
\includegraphics[width=9cm,height=6cm]{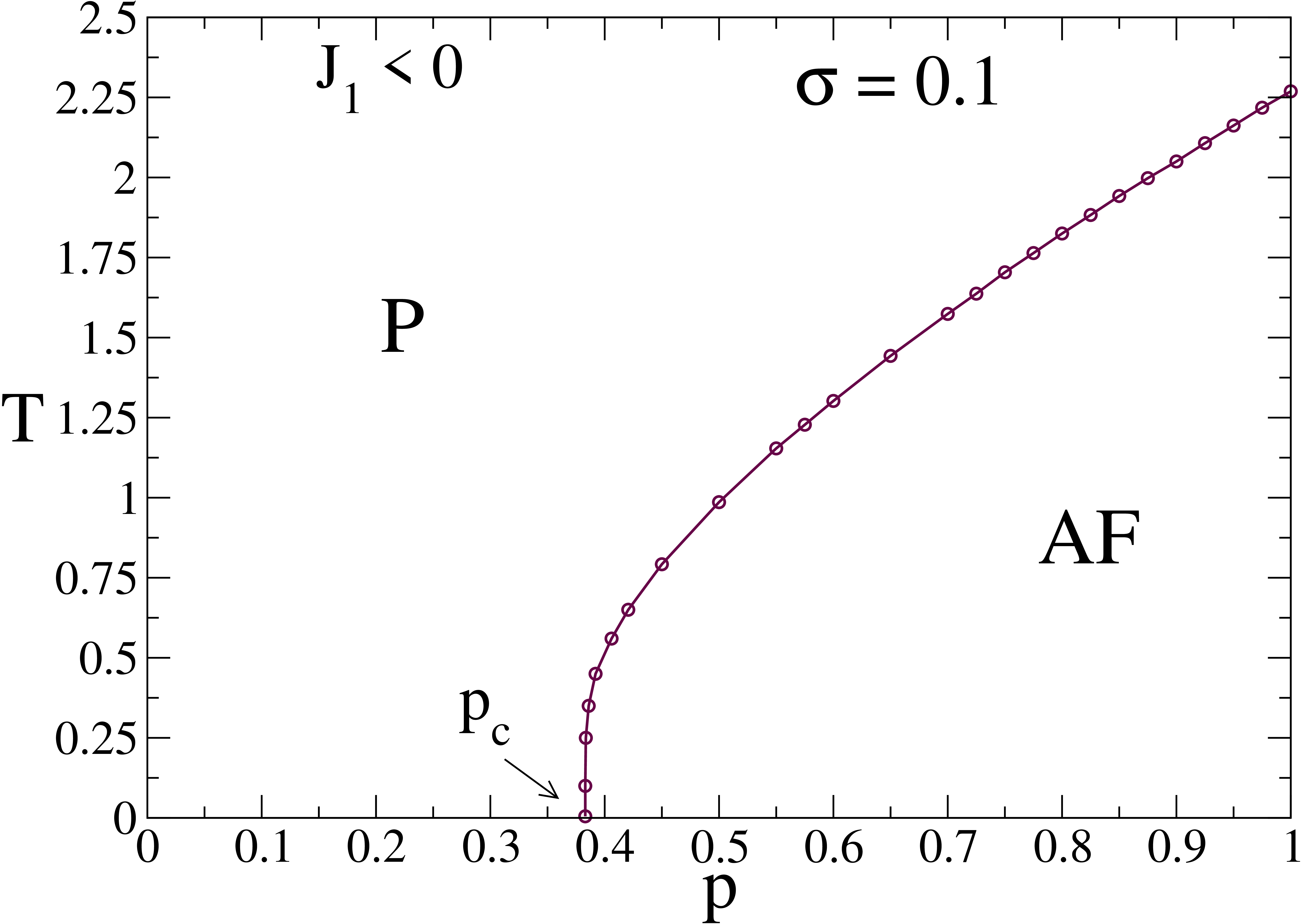}
\caption{\small{Phase diagram  for fixed antiferromagnetic $J_{1}$ bonds, and random $J_{2}$ interactions obeying the \textbf{PDF} given in Eq.(\ref{prob4}), when $\sigma = 0.1$. The horizontal axis is  the probability $p$, which is the weight of the first gaussian  in Eq.(\ref{prob4}). There is only  one frontier separating the \textbf{AF} and \textbf{P} phases. The \textbf{SAF} phase can not be formed for $\sigma \neq 0$.   } } 
\label{case1_df4}
\end{figure}
\vskip \baselineskip
\vskip \baselineskip

\begin{figure}[htp]
\includegraphics[width=7cm,height=5.5cm]{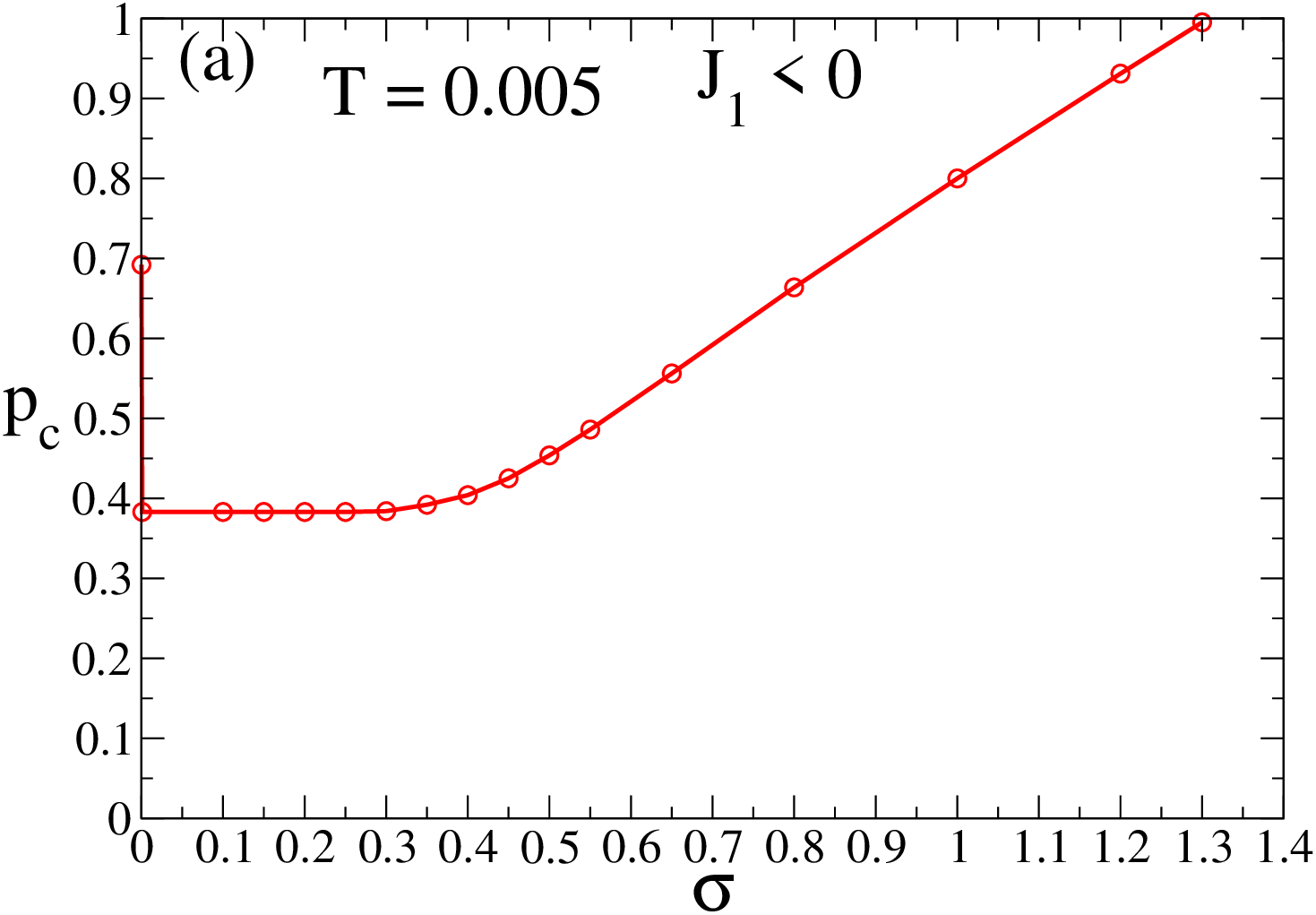}
\includegraphics[width=7cm,height=5.5cm]{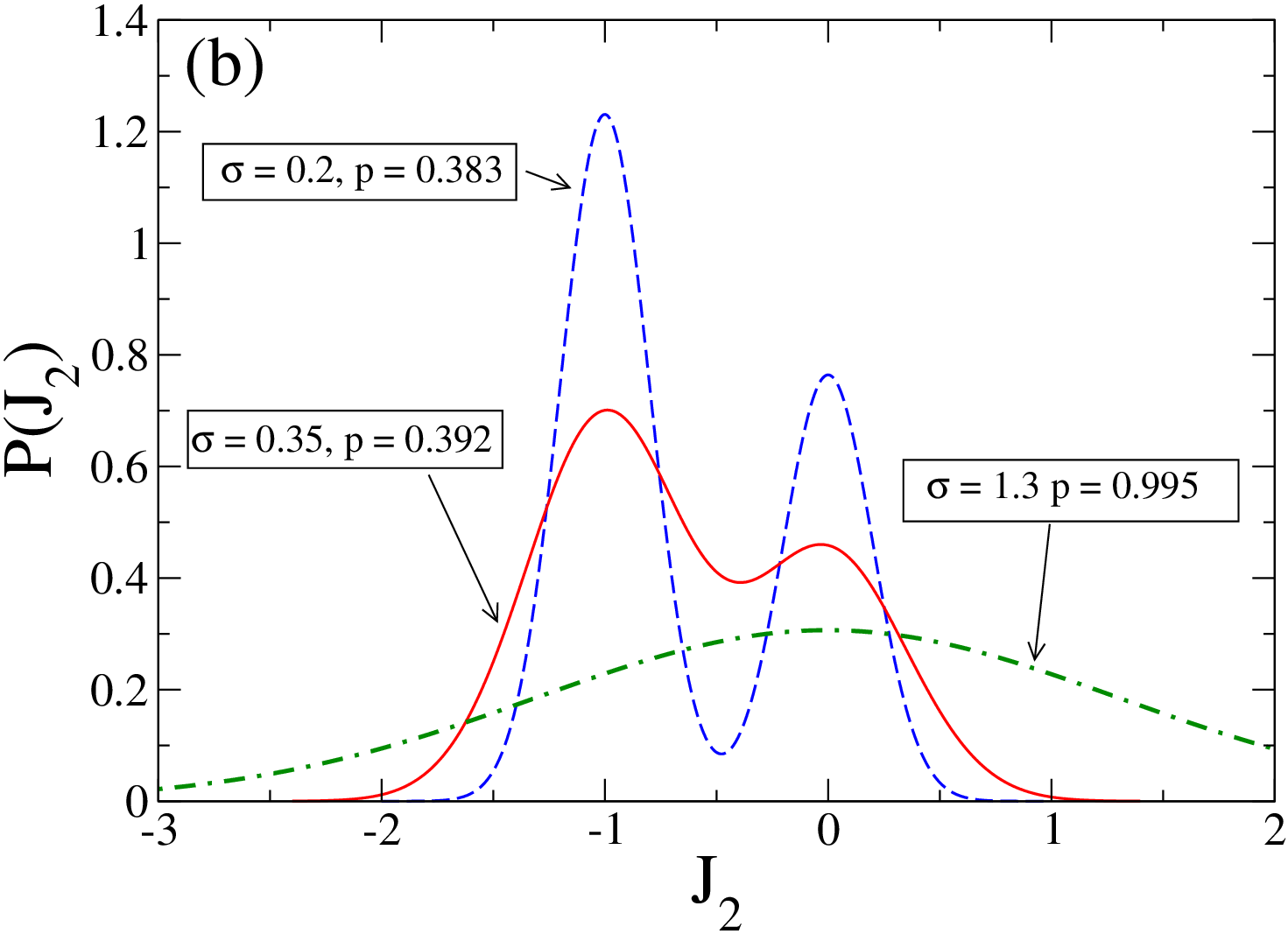}
\caption{\small{ (a) The critical probability of the phase diagram shown in Fig\ref{case1_df4} versus the parameter $\sigma$ given in Eq.(\ref{prob4}). Note the discontinuous fall in $p_{c}$ when $\sigma$ assumes a finite value.  In (b) is shown the \textbf{PDF} given in Eq.(\ref{prob4}) for three different values of the pair ($p$, $\sigma$).  } } 
\label{pcc1}
\end{figure}
\vskip \baselineskip
\begin{figure}[htp]
\includegraphics[width=7cm,height=5.5cm]{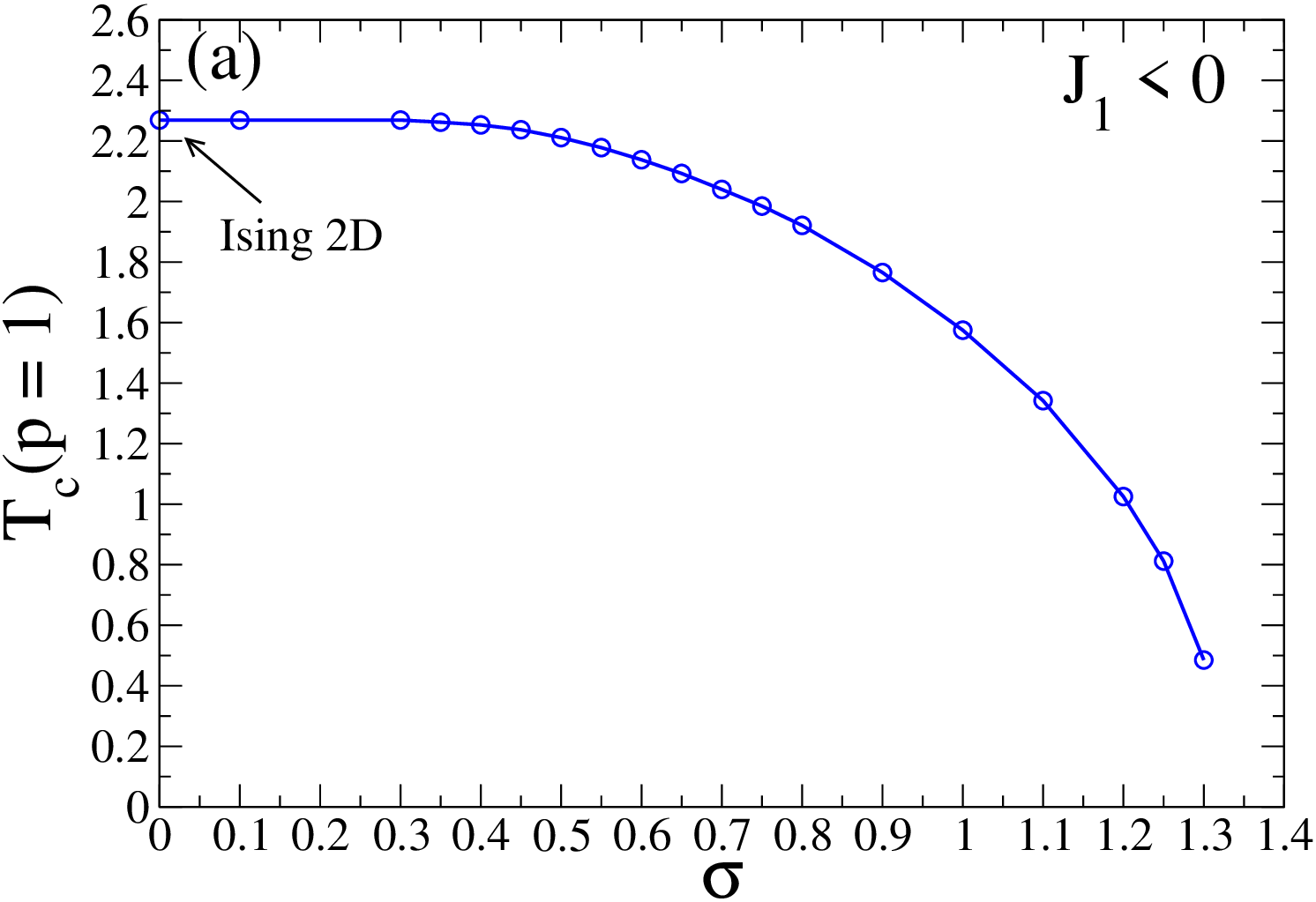}
\includegraphics[width=7cm,height=5.5cm]{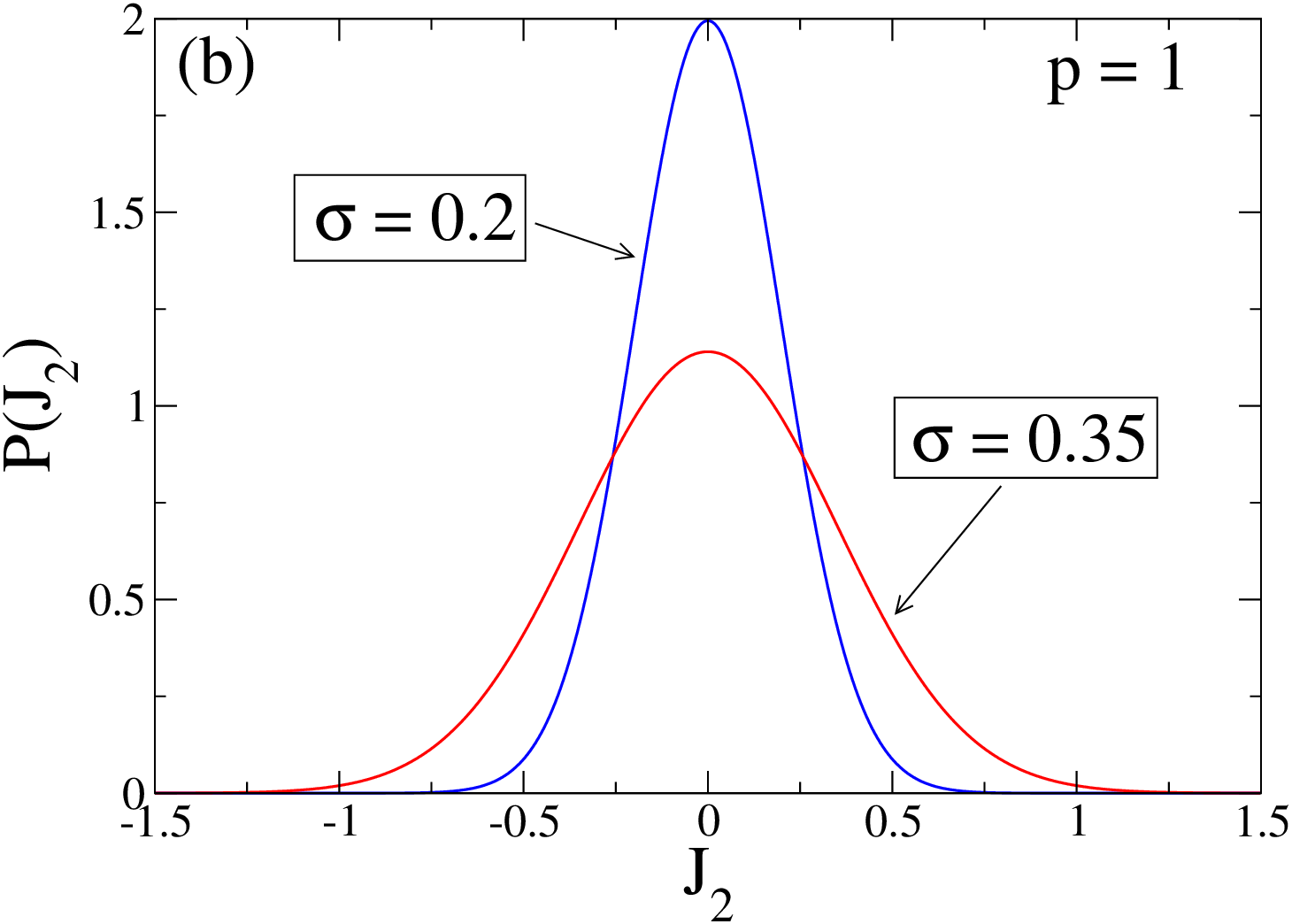}
\caption{\small{ (a) The critical temperature of the phase diagram shown in  Fig\ref{case1_df4} versus the parameter $\sigma$ given in Eq.(\ref{prob4}), for $p=1$. Note that $\sigma$ does not affect the value of the critical temperature for  $\sigma \lesssim 0.35$.  In (b) is shown the \textbf{PDF} given in Eq.(\ref{prob4}) for two different values of $\sigma$, for $p=1$. Note that the tails of the gaussian for $\sigma = 0.35$ include  $|J_{2}| > 1$, whereas for $\sigma = 0.2$ they die for  $|J_{2}| < 1$.   } } 
\label{tcc1}
\end{figure}
\vskip \baselineskip
\begin{figure}[htp]
\centering
\includegraphics[width=6.5cm,height=5.5cm]{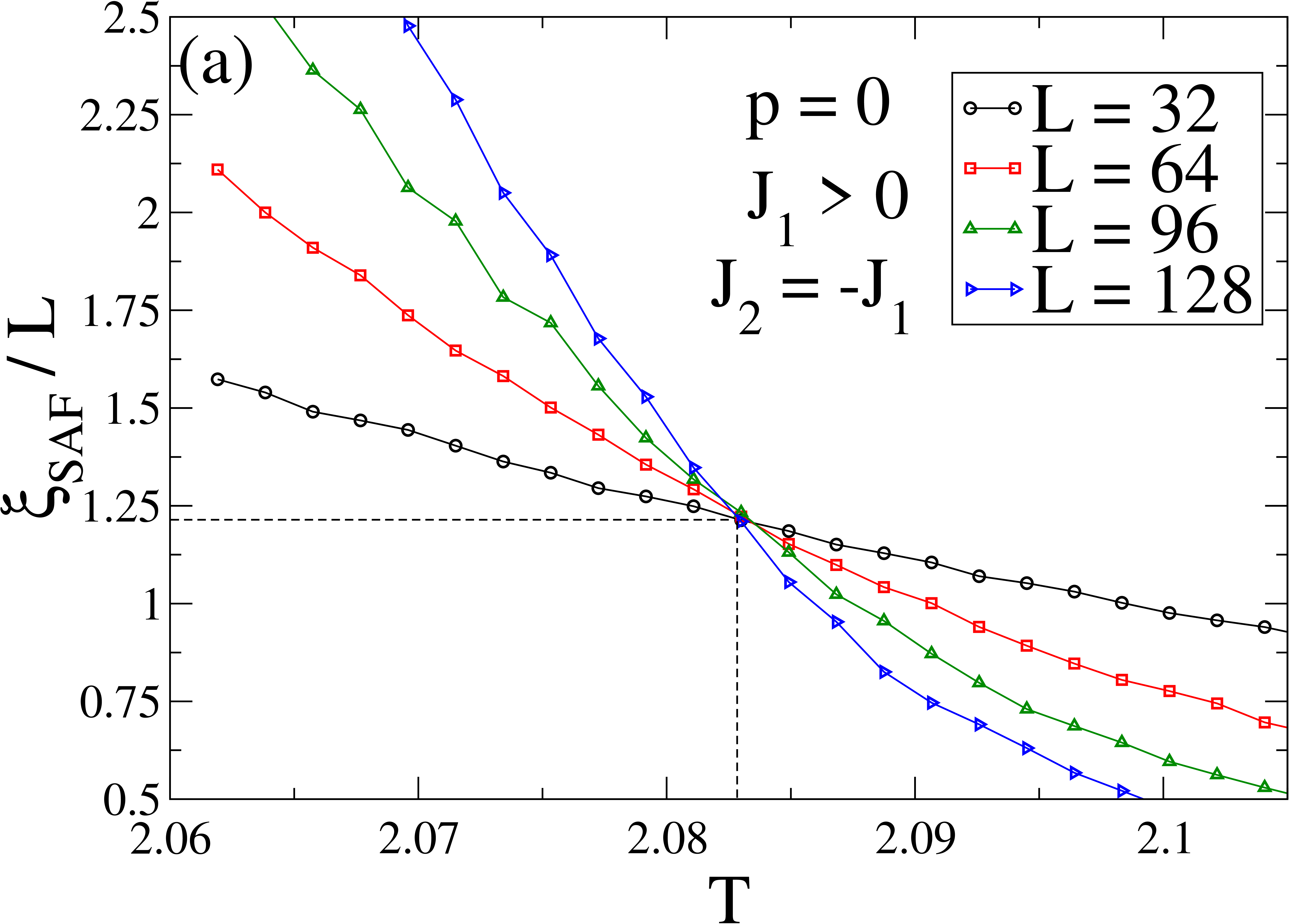}
\includegraphics[width=6.5cm,height=5.5cm]{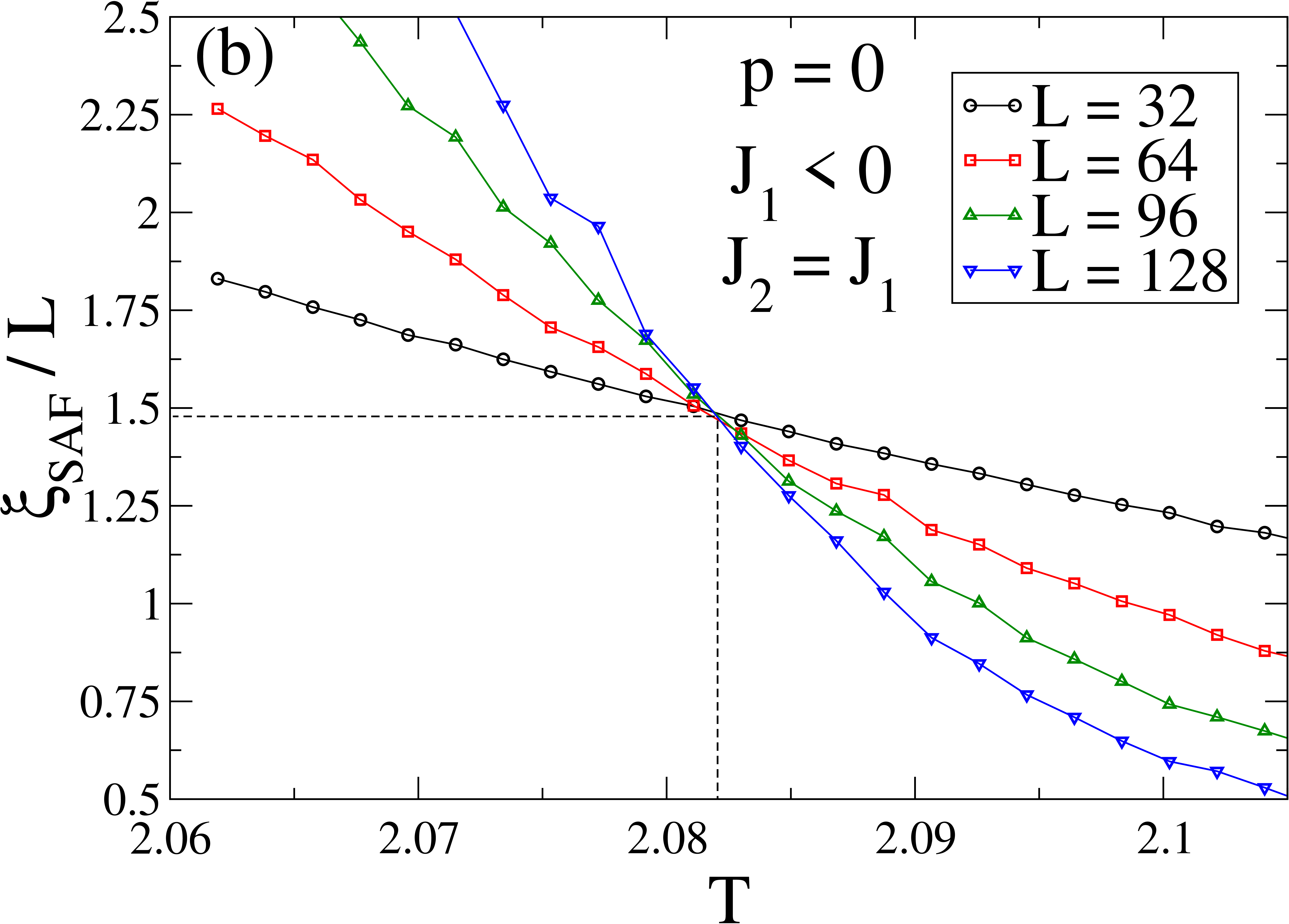}
\caption{\small{Curves of the correlation length  for different sizes of the lattice. Curves in (a) and (b) intercept at the critical temperature of the \textbf{SAF-P} transition  in Figures \ref{case1_df1} and \ref{case1_df3}, for $p=0$, respectively. In (a) the crossing point is observed at $(2.0832 \pm 0.0006, 1.210 \pm 0.005)$ ; In (b) it is at $(2.082 \pm 0.001, 1.483 \pm 0.008)$. Dashed lines are guides to the eye, through which we may note that  $x^{*}_{SAF}$ is different  in (a) and (b).} } 
\label{curves_p0J1J2}
\end{figure}
\vskip \baselineskip
\vskip \baselineskip
\begin{figure}[htp]
\centering
\includegraphics[width=9cm,height=6cm]{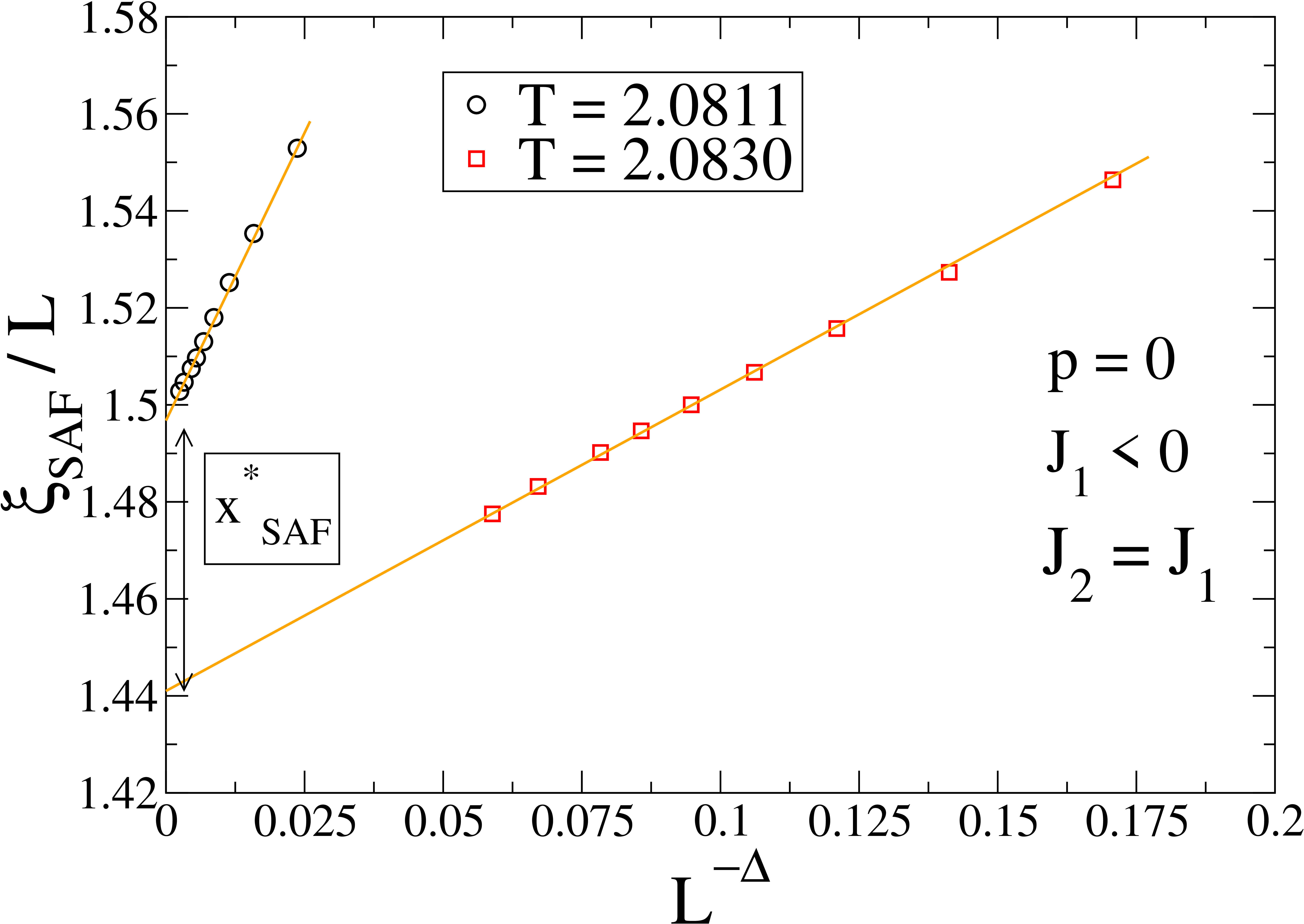}
\caption{\small{ Extrapolation  of $\xi_{SAF}/L$ to determine the amplitude $x^{*}_{SAF}$ for $L \to \infty$ (see the  intercepts at the vertical axis). It is based on Eq.(\ref{xiL})}. The lines correspond to the limits of the interval of the uncertainty of $T_{c}$ (read the caption of Fig.\ref{curves_p0J1J2}b). Therefore, the points between the intercepts are $1.440 < x^{*}_{SAF} < 1.497 $.  } 
\label{extrap1}
\end{figure}
\vskip \baselineskip
\vskip \baselineskip
\begin{figure}[htp]
\centering
\includegraphics[width=6.5cm,height=5.5cm]{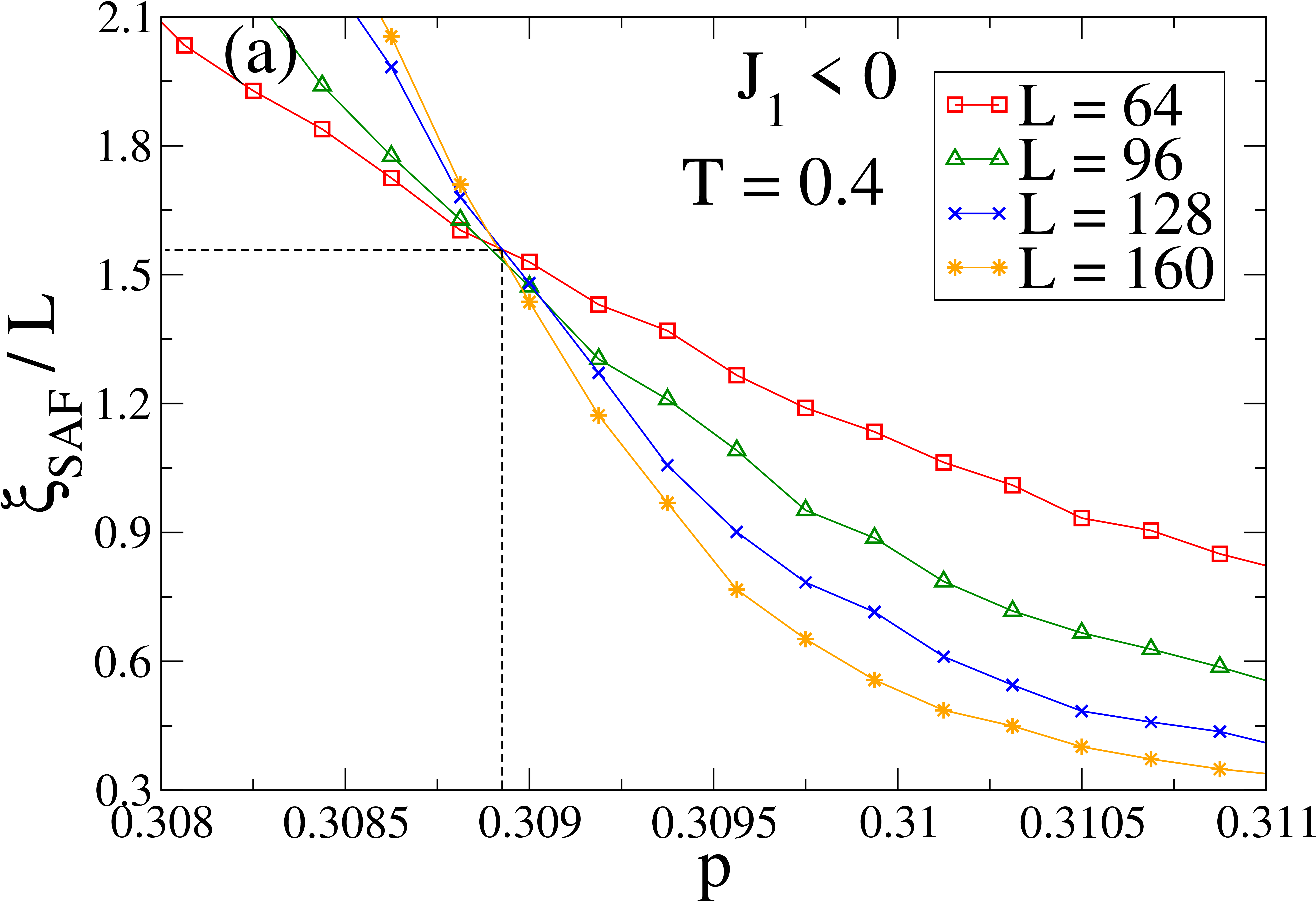}
\includegraphics[width=6.5cm,height=5.5cm]{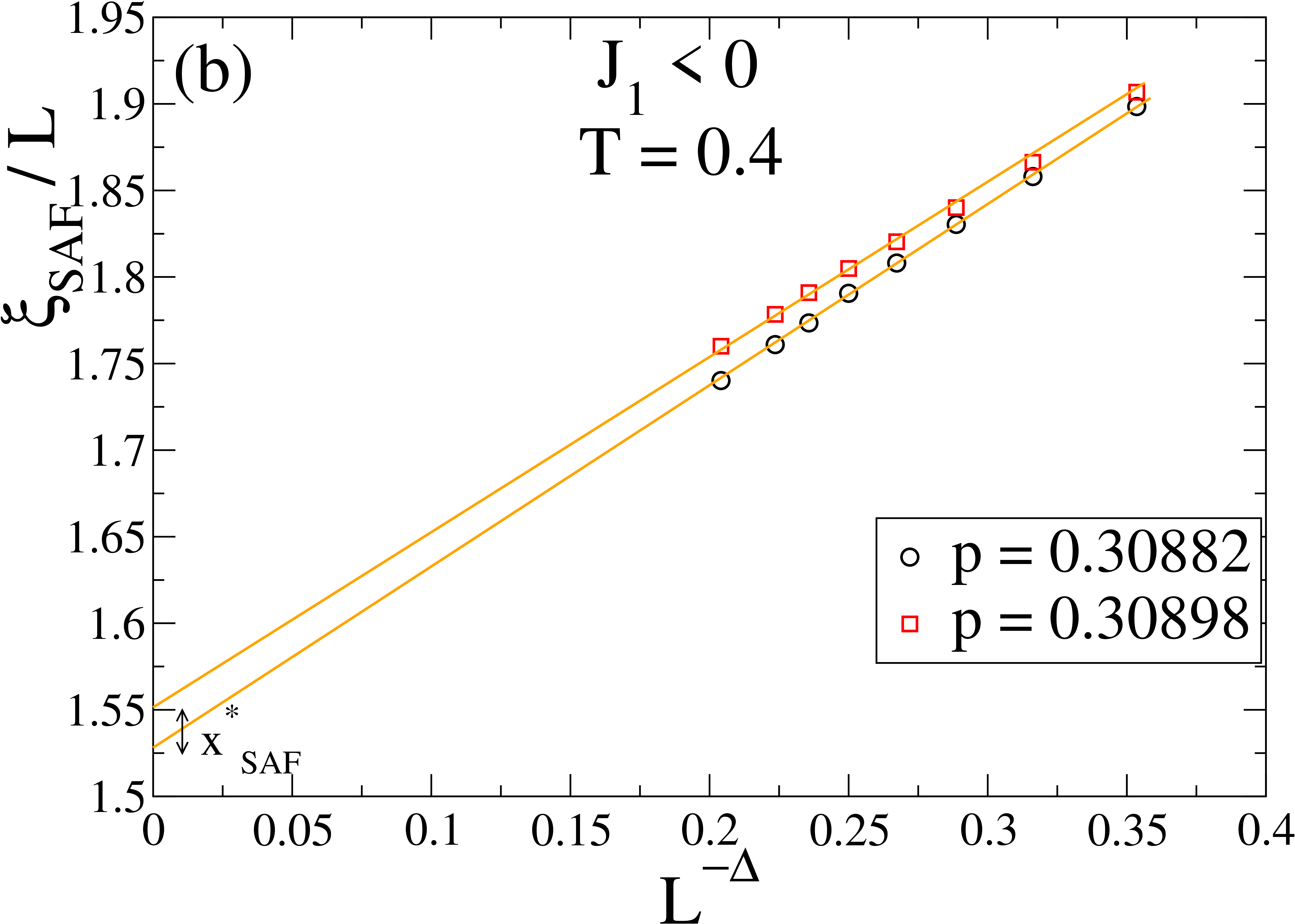}
\caption{\small{In (a) the uncertainty of the crossing point gives the critical probability $p$ of the \textbf{PDF} in Eq.(\ref{prob3}), in the interval $ 0.30882 < p_{c} < 0.30898 $. In (b) the limits of the uncertainty of $x^{*}_{SAF}$ are determined by extrapolating $\xi_{SAF}/L$ to $L \to \infty$, for each limit of $p_{c}$. So we have $1.53 < x^{*}_{SAF} < 1.55 $. This interval does not intersect  that of Fig.\ref{extrap1}.  } } 
\label{curves_T04}
\end{figure}
\vskip \baselineskip
\end{document}